\documentclass[journal,draftcls,onecolumn,12pt,twoside]{IEEEtranTCOM}

\usepackage{color}     
\usepackage{epsf,psfrag,amssymb,amsfonts,latexsym,cite,verbatim,enumerate,subfigure}
\usepackage{srcltx,amsmath,cases, graphicx}
\usepackage{times, multirow, array}
\usepackage[mathscr]{eucal}
\usepackage[normalem]{ulem}  
\usepackage[justification=centering]{caption}

\def\psfancypar#1#2{\begingroup\def\par{\endgraf\endgroup\lineskiplimit=0pt}
               \setbox2=\hbox{\large\sc #2}
               \newdimen\tmpht \tmpht \ht2 \advance\tmpht by \baselineskip
               \font\hhuge=Times-Bold at \tmpht
               \setbox1=\hbox{{\hhuge #1}}
               \count7=\tmpht \count8=\ht1
               \divide\count8 by 1000 \divide\count7 by \count8 
               \tmpht=.001\tmpht\multiply\tmpht by \count7 
               \font\hhuge=Times-Bold at \tmpht
               \setbox1=\hbox{{\hhuge #1}}
               \noindent
                \hangindent1.05\wd1
               \hangafter=-2 {\hskip-\hangindent
               \lower1\ht1\hbox{\raise1.0\ht2\copy1}%
                \kern-0\wd1}\copy2\lineskiplimit=-1000pt}

\newcommand{\E}{\mbox{{\rm E}}}

\def\boxit#1{\vbox{\hrule\hbox{\vrule\kern3pt
        \vbox{\kern3pt#1\kern3pt}\kern3pt\vrule}\hrule}}

\def\reals{ { {\rm  I \kern-0.15em R }  } }
\def\complex{ {\,{{\rm C} \kern-0.50em \raise0.20ex {  |}}\, }}

\def\mubf{\hbox{\boldmath$\mu$\unboldmath}}

\def\Sigmabf{\hbox{$\bf \Sigma$}}

\def\Thetabf{\hbox{$\bf \Theta$}}
\def\Lambdabf{\mbox{$ \bf \Lambda $}}

\def\Pibf{{\bf \Pi}}

\def\nbf{{\bf n}}

\def\qbf{{\bf q}}
\def\rbf{{\bf r}}
\def\sbf{{\bf s}}

\def\wbf{{\bf w}}
\def\xbf{{\bf x}}
\def\ybf{{\bf y}}
\def\zbf{{\bf z}}
\def\rbf{{\bf r}}
\def\xbf{{\bf x}}
\def\ybf{{\bf y}}
\def\Abf{{\bf A}}
\def\Bbf{{\bf B}}
\def\Cbf{{\bf C}}
\def\Dbf{{\bf D}}
\def\Ebf{{\bf E}}
\def\Fbf{{\bf F}}
\def\Gbf{{\bf G}}
\def\Hbf{{\bf H}}
\def\Ibf{{\bf I}}

\def\Kbf{{\bf K}}

\def\Mbf{{\bf M}}

\def\Pbf{{\bf P}}
\def\Qbf{{\bf Q}}
\def\Rbf{{\bf R}}

\def\Tbf{{\bf T}}
\def\Ubf{{\bf U}}
\def\Vbf{{\bf V}}
\def\Wbf{{\bf W}}
\def\Xbf{{\bf X}}

\def\Ec{{\cal E}}

\def\be{\vskip .3cm \begin{equation}}
\def\ee{\end{equation} \vskip .4cm \noindent}
\def\defeq{{\stackrel{\Delta}{=}}}
\newcommand{\R}{\mbox{$\hat {\bf R}_{N}$}}

\def\Rxx{\Rbf_{\ssstyle X\kern-.1em X}}

\let\ssstyle=\scriptscriptstyle

\def\Kout{\setbox1=\hbox{\Huge\bf K}\hbox to
1.05\wd1{\hspace{.05\wd1}
}}
\def\Sout{\setbox1=\hbox{\Huge\bf S}\hbox to 1.05\wd1{\hspace{.05\wd1}
}}

  \ifx\LabelFigloaded\MYundefined\relax
  \else
   \fi

  \def\LabelFigloaded{\relax}

  \chardef\LabelFigCatAt\the\catcode`\@
  \catcode`\@=11

 \let\LabelFigwlog@ld\wlog
 \def\wlog#1{\relax}

 \ifx\\\MYundefined@
    \let\\\relax
 \fi

  \def\ms@g{\immediate\write16}

 \def\N@wif{\csname newif\endcsname }
 \def\Temp@ {\N@wif\ifIN@}
 \ifx\INN@\MYundefined@
    \else \let\Temp@\relax
 \fi
 \Temp@

  \def\IN@{\expandafter\INN@\expandafter}
  \long\def\INN@0#1@#2@{\long\def\NI@##1#1##2##3\ENDNI@
    {\ifx\m@rker##2\IN@false\else\IN@true\fi}%
     \expandafter\NI@#2@@#1\m@rker\ENDNI@}
  \def\m@rker{\m@@rker}
 
  \newtoks\Initialtoks@  \newtoks\Terminaltoks@
  \def\SPLIT@{\expandafter\SPLITT@\expandafter}
  \def\SPLITT@0#1@#2@{\def\TTILPS@##1#1##2@{%
     \Initialtoks@{##1}\Terminaltoks@{##2}}\expandafter\TTILPS@#2@}

 \def\Shifted@@#1#2#3{\setbox0=\hbox{#3}%
   \raise -\dp0\vbox {\kern-#2%
       \hbox {\kern#1\unhbox0\kern-#1}%
           \kern#2}}

 \newcount\gridcount
 \newbox\auxGridbox@ \newbox\hGridbox@ \newbox\vGridbox@
 \newbox\Labelbox@ \newbox\auxLabelbox@
 \newbox\Coordinatebox@
 \newtoks\Labeltoks@
 \newdimen\Wdd@ \newdimen\Htt@
 \newdimen\Wddd@ \newdimen\Httt@
 
 \def\Wr@{\immediate\write16}

 \newdimen\GL@wd
 \def\GridLineWidth#1{\GL@wd=#1}

 \def\gobble#1{}
 \def\EdgeErr@{\Wr@{}%
      \Wr@{\string\Edges\space argument
      1, 10, 100 or 1000 please\string!}%
      }

 \newcount\Edgect@

 \def\Sweepup#1\endSweepup{}

 \def\SetEdges@{%
    \edef\Zr@@s{\expandafter\gobble\number\Edgect@\empty}%
        \count255=0\Zr@@s\relax
        \ifnum\count255=\z@\else\EdgeErr@\show\tailtest\fi
        \count255=1\Zr@@s\relax
        \ifnum\count255=\Edgect@\relax\else\EdgeErr@\show\leadtest\fi
    \EdgGl@b\edef\Zr@s{\expandafter\gobble\Zr@@s\empty}
    \ifnum\Edgect@>\@ne\relax\EdgGl@b\let\L@Dc\empty
        \else\EdgGl@b\edef\L@Dc{\string.}\fi
    \ifnum\Edgect@>\@ne\relax
        \EdgGl@b\edef\Edgescale@##1{\divide##1 by \Edgect@}%
        \else\EdgGl@b\edef\Edgescale@##1{}\fi
    }

 \def\Edges#1{\Edgect@=#1\relax
     \let\EdgGl@b\global \SetEdges@}

 \Edges{1}

 \def\hhrule{\hrule height \GL@wd\vskip-.\GL@wd}

 \def\hRule@{%
   \advance\gridcount -2%
   \vfil\hhrule\vfil
   \llap{\smash{\raise -2.5pt
     \hbox{\L@Dc\number\gridcount\Zr@s\kern2pt}}}%
   \hhrule
   }

\def\vvrule{\vrule width \GL@wd \kern-\GL@wd}

 \def\vRule@{\advance\gridcount 2%
   \hfil\vvrule\hfil
   \setbox\auxGridbox@=\vbox to 0pt
      {\vskip \Htt@\vskip 2pt
        \hbox to 0pt{\hss\L@Dc\number\gridcount\Zr@s\hss}\vss}%
      \wd\auxGridbox@=0pt \box\auxGridbox@
   \vvrule
   }

 \def\PlaceGrid@@{\gridcount=10 
  \setbox\hGridbox@=\hbox{%
        \hbox{%
             \hskip-.4pt\vrule
             \vbox to \Htt@{%
               \offinterlineskip\parindent=\z@\relax
               \hbox to \Wdd@{\hfil}
               \hRule@\hRule@\hRule@\hRule@
               \vfil\hhrule\vfil}%
             \vrule\hskip-.4pt}
    }%
  \gridcount=0%
  \setbox\vGridbox@=\hbox{%
      \vbox{\offinterlineskip\parindent=0pt\hsize=0pt
         \vskip-.4pt\hrule%
         \hbox to \Wdd@{%
                 \vtop to \Htt@{\vfil}%
                 \vRule@\vRule@\vRule@\vRule@
                 \hfil\vvrule\hfil}%
         \hrule\vskip-.4pt}}%
  \wd\hGridbox@=0pt\ht\hGridbox@=0pt
  \wd\vGridbox@=0pt\ht\vGridbox@=0pt
  \hbox{\box\hGridbox@\box\vGridbox@}%
  }

 \def\LabelsGlobal{\def\LabGl@b{\global}}
 \def\LabelsLocal{\def\LabGl@b{}}
 \LabelsGlobal 

 \def\SetLabels#1\endSetLabels{%
   \LabGl@b\Labeltoks@={#1()\\}%
   }

 \LabGl@b\Labeltoks@={()\\}

 \def\ShowGrid{\LabGl@b\let\PlaceGrid@\PlaceGrid@@}
 \def\HideGrid{\LabGl@b\let\PlaceGrid@\relax}
 \def\Grids{\ShowGrid\LabGl@b\let\GridSwitch@\ShowGrid}
 \def\noGrids{\HideGrid\LabGl@b\let\GridSwitch@\HideGrid}

 \noGrids

 \def\bAdjust@@{%
     \setbox\auxLabelbox@=\hbox{\raise \dp\auxLabelbox@
            \box\auxLabelbox@}}
 \def\bAdjust@{\let\vAdjust@\bAdjust@@}

 \def\eAdjust@@{\dimen0=-.5\ht\auxLabelbox@
     \advance\dimen0 by .5\dp\auxLabelbox@
     \setbox\auxLabelbox@=
            \hbox{\raise\dimen0\box\auxLabelbox@}}
 \def\eAdjust@{\let\vAdjust@\eAdjust@@}

 \def\tAdjust@@{%
     \setbox\auxLabelbox@=\hbox{\raise-\ht\auxLabelbox@
            \box\auxLabelbox@}}
 \def\tAdjust@{\let\vAdjust@\tAdjust@@}

 \let\vAdjust@\relax

 \def\lAdjust@{\let\hAdjust@\rlap}
 \def\rAdjust@{\let\hAdjust@\llap}

 \let\hAdjust@\relax\let\vAdjust@\relax

 \def\FetchLabel@#1(#2)#3\\{%
     \IN@0#2@@\ifIN@
        \setbox0=\hbox{\ignorespaces#1#3\unskip}%
        \ifdim\wd0>0pt
           \ms@g{}%
           \ms@g{ !!! Bad label(s)? !!!}%
           \message{ #1(#2)#3}%
        \fi
        \def\LabelMole@##1\endFetchLabel@{%
            \IN@0()\\@##1@%
            \ifIN@\def\Temp@{\FetchLabel@##1\endFetchLabel@}%
            \else\def\Temp@{}%
            \fi
            \Temp@
           }%
     \else
       \ignorespaces#1\unskip
       \setbox\auxLabelbox@=%
         \hbox to 0pt{\hss\ignorespaces\hAdjust@
          {\ignorespaces#3\unskip}\hss}%
       \vAdjust@
       \let\hAdjust@\relax\let\vAdjust@\relax
       \AugmentLabelBox@@{#2}%
       \ht\Labelbox@=0pt\dp\Labelbox@=0pt
       \let\LabelMole@\FetchLabel@%
     \fi\LabelMole@}

 \newtoks\XYSep@ 
 \def\SetXYSeparator#1{%
     \IN@0#1@@\ifIN@\XYSep@{*}%
     \else
     \XYSep@{#1}%
     \fi
     }

 \SetXYSeparator*

 \def\AugmentLabelBox@@#1{%
     \IN@0\the\XYSep@ @#1@\ifIN@
       \SPLIT@0\the\XYSep@ @#1@%
       \setbox\Labelbox@=\hbox to 0pt{%
         \unhbox\Labelbox@
         \Shifted@@{\the\Initialtoks@\Wddd@}%
         {\the\Terminaltoks@\Httt@}%
         {\box\auxLabelbox@}}%
     \else
         \ms@g{}%
         \ms@g{ !!! Bad insertion point. !!!}%
         \message{ (#1\ this point was rejected.)}%
     \fi
    }

 \def\FetchOption@#1[#2]#3\endFetchOption@{%
    \def\temp{#1}
    \ifx\temp\empty
       \Edgect@=#2\relax
       \let\EdgGl@b\relax
       \SetEdges@
       \Cleaner@#3%
    \fi}

 \def\Cleaner@#1[@]{\Labeltoks@{#1}}
     
 \def\PlaceLabels@@{\mathsurround=0pt
     \def\Cr@{\\}%
     \let\L\lAdjust@\let\R\rAdjust@
     \let\B\bAdjust@\let\E\eAdjust@\let\T\tAdjust@
     \expandafter\FetchOption@\the\Labeltoks@[@]\endFetchOption@
     \Wddd@=\Wdd@ \Edgescale@\Wddd@ 
     \Httt@=\Htt@ \Edgescale@\Httt@
     \expandafter\FetchLabel@\the\Labeltoks@\endFetchLabel@
     \box\Labelbox@
     }%

 \let \PlaceLabels@\PlaceLabels@@

 \def\AffixLabels#1{\setbox\Coordinatebox@=\hbox{#1}%
      \Wdd@=\wd\Coordinatebox@ \Htt@=\ht\Coordinatebox@
      \advance\Htt@ \dp\Coordinatebox@
      \hbox{\copy\Coordinatebox@\kern-\Wdd@ 
           \Shifted@@{0pt}{-\dp\Coordinatebox@}%
           {\PlaceLabels@\PlaceGrid@}%
           \kern\Wdd@}%
      \GridSwitch@ 
      \LabGl@b\Labeltoks@{()\\}%
      }
 
   \let\wlog\LabelFigwlog@ld   
   \catcode`\@=\LabelFigCatAt  

                                By

              Raymond S\'eroul <A18645@FRCCSC21.BITNET>
                                and 
              Laurent Siebenmann <lcs@topo.math.u-psud.fr>
    
              VERSIONS: July 1991, Oct 1991, Jan 1992, July 1992

INTRODUCTION

      This labelling package is intended for TeX users who
rely on non-TeX sources for for their graphics inserts.  It
provides means for adding TeX labels to such inserts with a
minimum of fuss. 

       For most labels, TeX users have in the past found it
reasonably convenient to rely on non-TeX sources. Typical
occasions when an inescapable need for TeX labels seemed to
arise are

 (a) when the graphics program lacks certain exotic or complex
mathematical symbols

 (b) when the very highest typographical quality is wanted for the
labels

 (c) when labels included with the graphics fail to print, 
 and you cannot figure out why (cf. boxedeps.doc).  The labels
 provided by labelfig.tex are 100

       Since this package first appeared, many users, who in the
past scarcely dreamed of using TeX labels, have come to use
nothing but.  So it is now appropriate to add

Intoxication Warning:  TeX labels may be addictive and expensive. 

     If you have a fast preview you may disagree, and even find
that this package provides an agreeable paste-up environment; see
extra applications at end.

     Note to publishers: It is possible and convenient to ultimately
export the TeX labels produced by labelfig.tex to become an integral
part of the EPS file. This is often desired by a publisher who typically
uses an "upmarket" graphics or page layout program, with which the
staff is skilled in perfecting figures.  See Appendix I for
a recipe.

     The authors are grateful to Patrick Ion of Math Reviews for
helpful comments and encouragement.

BASIC INSTRUCTIONS

    After reading in the macro file using

preview or proof your figure with a coordinate grid printed on
top, by typing the following:

    \ShowGrid  
    \AffixLabels{<the graphics insertion>}

Here <the graphics insertion> is what you would type to insert
the graphics object alone without the grid.  This must provide
for the space around it. For example <the graphics insertion>
might well be \BoxedEPSF{MyFigure scaled 700} using the
boxedeps.tex macro package (from same source); this provides a
TeX box containing the encapsulated PostScript insert specified by
the file MyFigure. \AffixLabels{...} provides the grid (supposing
\ShowGrid is present) and later, once you have specified labels
using the grid, it will "tack on" the labels.

     The grid is a sort of (usually elongated) checkerboard of
ten rows and ten columns and its (internal) partitions are by
default numbered  .1, ... ,.9  both horizontally (X-coordinate
running left to right) and vertically (Y-coordinate running bottom
to top).  Thus the points enclosed by the grid correspond to the
points of the unit square in the cartesian "X-Y" plane, the lower
left corner corresponding to the origin (0,0).  By extrapolation,
the full page corresponds to a larger rectangle in the plane.

     These coordinates serve to position labels as follows.
Before the \AffixLabels{...} command type label specifications:

  \SetLabels
   (<X-coordinate>*<Y-coordinate>) <first label> \\
   .
   .
   .
   (<X-coordinate>*<Y-coordinate>)  <last label> \\
  \endSetLabels

Each row specifies one label and is terminated by \\.  In each
row, the position indicator comes first; it is written as a
standard cartesian point except that the X- and Y- coordinates
are separated by * rather than a comma because TeX allows a
comma as decimal point. There are no dimension units to specify
as the unit is the grid itself.

     By default, this cartesian point specifies where the middle
of the baseline of the label will be located.  However if you precede
the point by \L [or \R] the left [or right] edge of the baseline will
be located there. Similarly you may also precede the point by \T, \E,
or \B to vertically align the top equator or bottom of the label box
at the specified point.  This gives nine standard positions of
the label with respect to the insertion point --- corresponding to
the eight principle points of the compas and the center

                     \L\T     \T      \R\T

                     \L\E     \E      \R\E

                     \L\B     \B      \R\B

But this neglects the default "baseline" level of TeX,
giving potentially three more positions

                     \L    <no tag>   \R

For text, the baseline level is often the preferred. Its relation to
the others is variable. It will often coincide with the bottom level,
as happens for "X".  But it is often distinct, as for "g", in which
case you have in all 12 distinct positions rather than 9.

     It is convenient to think of this specification of label
position as attaching the label by a thumb-tack to the coordinate
grid. There are up to twelve positions of the thumb-tack on the
label, while the position of the thumb-tack on the coordinate grid is
arbitrary.  Normally, one choses the position of the thumb-tack on
the label to be the one that is the closest to the item being
labeled.  There are good reasons for this "rule of thumb":

   (a)  It facilitates correct positioning at first try.

   (b)  If the scale of the figure must be altered after labels
have been affixed, the labels have a good chance of remaining well
positioned.

   (c)  The visible grid need not extend beyond the "bounding box"
for the figure, because the best preferred position is always
(at least almost) within the bounding box .

The second reason is particularly important. Indeed it often
happens that scale has to be altered after labelling begins, in
order to either provide space for the labels, or to adjust
proportions between the labels and the figure.  (The size of labels
is unaffected by scaling.)

     Here is an artificial but self-contained test which uses
TeX rules to make a graphics object.

TEST

    Do not skip this!

 \def\FrameIt#1{\hbox{\vrule$\vcenter {\hrule\kern3pt%
             \hbox {\kern3pt #1\kern3pt}%
               \kern3pt\hrule}$\relax\vrule}}

 \def\Caption#1#2{\FrameIt{%
       \vtop {\hsize=#1\relax \parindent=0pt
         \leftskip=0pt \rightskip=0pt plus15pt
         \parfillskip=0pt
         \lineskip=1pt\baselineskip=0pt
         #2}}}

 \def\FirstQuadrant{\hbox to 100pt{\vrule\vbox to 100pt{%
        \hbox to 100pt{\hfil}\vfil\hrule}\hss}}

  \SetLabels
    \R(.5*.2) $\zeta\,\cdot$\\
    (.9*-.10) $\xi$\\
    \R(-.03*.9) $\eta$\\
    \T(.5*.9) \Caption{70pt}{%
          \it The norm of
          $g(\xi+i\eta)$ is indicated on
          contours of this invisible surface.}\\
  \endSetLabels

  \AffixLabels{\FirstQuadrant}

  \end

  Note that the coordinates to use for labels are indicated on the
edges of the grid (when visible) corresponding to the conventional
x- and y- axes of the Cartesian plane. By default the grid is
1-by-1. However, by the command \Edges{100}, you can change this
to 100-by-100 and many users find this alternative most
convenient. Place the command \Edges{...} in your style file (or
header) since its effect is is global. Other possible edge values
are 10 and 1000.

  If you use the command \Edges{...} at all, do so with care.  For
if you accidentally delete an \Edges{...} command your labels will
abruptly be badly misplaced and may logically but mysteriously
generate "dimension too big" errors under TeX and "off page" errors
under your driver.  

  You can dictate the edgescale for an individual figure by giving
the scale in brackets immediately after \SetLabels.  Thus, to
import into an article using say \Edge{100} a figure labelled using
another edgescale, say the original 1-by-1 default, you can use
\SetLabels[1]...\endSetLabels.

GETTING IT DOWN PAT

     Complicated labeling deserves the same respect as
complicated mathematics.  Do not expect it to come out perfect the
first time!  What is needed in either case is a mechanism to
repeatedly typeset troublesome pieces.

     One mechanism is always available.  One does complicated
labelling in a separate "test" file involving just the figure being
labelled;  a texpert will know how to \dump TeX's current state as
a temporary format that restarts rapidly at each retry.  Usually,
one then pastes the completed labelled figure back into the main
TeX file, but, of course, one can also \input it as an auxiliary
file.

     If you do not have a TeXpert at handy, here is a first
approximation to an efficient setup. By deletions reduce a copy
of your article to just a few lines before and after the figure.
Now label the figure, and finally, copy and paste the labelled
figure to the original article. Then copy the next figure to label
into this testbed and repeat. The TeXpert can improve the  speed
at which TeX starts up, by compiling a format specifically for
your article; just one caution: best NOT include in the format
ephemeral details of setup like \Set<mydriver>ArtSpecials (from
boxedeps.tex because this reads  figure dimensions which you may
change during your work session.

     An improved mechanism to repeatedly typeset troublesome
pieces is now available on the Macintosh; it is called LinoTeX;
see the same ftp sources.  It could be set up on many types
of computer.

     Before using labelfig.tex to attach labels to a graphics
object inserted using boxedeps.tex or BoxedArt.tex, make it a
firm rule to carefully adjust the bounding box using the trimming
commands of these packages, and also at least tentatively scale
and position the object. Beware of changing the grid inadvertently
after the labels have been positioned.  For example, correcting
the bounding box of a PostScript graphics object can foul up the
labels by changing the coordinate grid to which the labels are
attached. This is particularly true for the trimming  commands of
boxedeps.tex and BoxedArt.tex. However, as noted already, change
of scale is much less disruptive, and modest adjustments should be
well tolerated.

     Sometimes the labels protrude so far from the bounding box
of a figure that the figure has to be repositioned.  Best do this
by ad hoc spacing, say using \hglue and \vglue; altering the
bounding box would create a vicious circle.

     Remember that you are responsible for preventing labels
from overlapping. You are responsible for all label typography
including size and style. A label is really just about anything
that can be put in a TeX box. Note that spaces at the beginning
and end of labels will normally be suppressed; if you really want
them you must protect them with TeX braces.

     This package temporarily sets the \mathsurround parameter
of TeX to zero  while the labels are being affixed. This is done
because nonzero \mathsurround space would influence the position
of left and right aligned labels; then, when a texpert or printer
modifies mathsurround, diagram labeling might be disastrously
altered. There is a small price to pay involving labels that are
formatted as caption boxes including mathematics: you  may want or
need to specify an explicit mathsurround space within the caption
box; it will not influence anything outside.

     Those hostile to the use of * as separator between
the X and Y coordinates of label insertion points, are free to
impose another using \SetXYSeparator{<the new separator>}.  
Americans may prefer "," to "*" since they never use a 
comma as a decimal point; on the other hand, * may be more visible.

APPENDIX (I)  MERGING labelfig.tex LABELS INTO AN EPSF GRAPHICS OBJECT.

     As promised in the introduction, here is a recipe useful for
publishers. It works at least on Macintosh and at least for vectorized
graphics and Adobe type1 fonts.  (There is surely a similar recipe for
PCs under MSWindows.)

 (a)  Use boxedeps.tex utility to integrate the figure given by the eps
file, "x.eps" say, with a visible frame around it.  See
\ShowDisplacementBoxes command in boxedeps.tex.  To get precise results
automatically it is important to use the \Trim... commands of
boxedeps.tex making the "DisplacementBox" neatly fit the figure.

 (b)  Use the TeX printer driver and LaserWriter (versions >= 8.1.1) to
export to an EPSF the DVI page containing the integrated, labelled
figure. You now have an EPS file  "xx.eps"  that contains too much, and at
the wrong scale, and at wrong position.

 (c)  Convert the EPSF to an Adode Illustrator format EPSF using
the shareware utility called epsConvert by Sam Weiss
1993-- (currently $25).

 (d)  In Illustrator (or a compatible program), group the labels and the
"DisplacementBox"; copy them to the clipboard and paste them into "x.ps".
This step requires that all the label fonts be "visible to the Macintosh.

 (e)  Translate and scale the pasted group consisting of the labels plus
the "DisplacementBox" so as to make the "DisplacementBox" the bounding
box of (labelless) figure represented by "x.eps".  At this point the
labels will be correctly placed on the figure "x.eps".

 (f)  Ungroup and delete the "DisplacementBox".  The result is the
desired single EPS file, "x+.eps" say, It contains the original figure
plus its labels.  

     Using grouping and ungrouping appropriately in "x+.eps", a
publisher's staff can very efficiently improve label positions etc.

APPENDIX II)  SOME EXOTIC APPLICATIONS

     The grid of labelfig.tex is analogous to a light-table in
classical page makeup with wax or latex glue.  In principle, you
can use it to compose any page from its indivisible parts.  This
even has some of the artisanal charm of classical paste-up
provided you have a fast screen preview to make the process
"interactive".

     In practice labelfig.tex is a tool for nonstandard jobs.
Here are a few going beyond the labelling already discussed.

(I)  GRAPHICS INTEGRATION.

     This is accomplished by treating the imported graphics
objects as labels.  The underlying graphics object is then
typically an empty  \vbox to <dimension>{\vfill} in a TeX
\midinsert...\endinsert construction.  A label line
might be of the form

   (.1*.1) \\

The exact form of the special command varies from driver to
driver.  However, in the case of encapsulated PostScript graphics
(EPSF norm), by relying on boxedeps.tex, one can have the
following standard syntax (independant of driver  (see
boxedeps.doc for details.
  
  (.1*.1) \BoxedEPSF{MyFigure scaled <scale in mils>}\\

This may be slow since it requires TeX to read the PostScript
file to read bounding box using many complex macros.  So you
may want to try

  (.1*.1) \EPSFSpecial{MyFigure}{<scale in mils>}\\

which is fast and driver independant, but it squashes the
bounding box, normally to its lower left corner.

     Similarly for graphics of the Macintosh PICT norm ---
using BoxedArt.tex (same sources) in place of boxedeps.tex.

     This approach to integration is to be recommended when
one is assembling a composite graphics object.

 (II)  COMMUTATIVE DIAGRAM ENHANCEMENT

     Commutative diagrams or arrays of mathematical objects
connected by arrows of various sorts are common in mathematics.
The mathematical objects require the use of TeX.  Recently TeX
acquired a good collection of arrows of all slopes --- that of
LamSTeX --- plus pwerful macros to build the diagrams.

     However, even the LamSTeX collection is often
inadequate; it lacks for example double shafted arrows, dotted
arrows and curved arrows. Fortunately it is possible to produce
such arrows on an individual basis using sophisticated graphics
programs such as Illustrator and AldusFreehand (both serving
the EPSF norm) or using Metafont (with its public domain norm).
Since the creation of each new arrow is a work of love, you
probably want to limit the number of arrows by using LamSTeX
for most arrows. The 40K commutative diagram module of LamSTeX
has been adapted to work with AmSTeX and a copy may be posted
with LabelFig and related files. Unfortunately no one has yet
offered a version that works with Plain TeX or LaTeX.

       Suffice it here to say that when the exotic arrow has
been somehow imported into TeX, labelfig.tex treats it as a
label that one affixes to the commutative diagram.  Two other
steps will be treated in separate notes, namely the matter of
extracting the dimension specifications for the arrow and the
construction of the arrow --- for these steps are far from
unique and often depend intimately on your computer environment. 
Notes for the Macintosh-Textures-Illustrator combination are
found in the file ExoticArrows.doc.

 (III) NESTING 

Ingenuity pays off in exploiting labelfig.tex. One can
mix graphics and typography quite freely.  labelfig.tex is good
for freeform or overlapping arrangements, while boxedeps.tex (or
BoxedArt.tex) is best for regimented non-overlapping
arrangements --- and the two can be combined.

     The default behavior of labelfig.tex is not ideal 
for nesting objects, because to prevent trouble for beginners
the register for labels is globally cleared when \AffixLabels
concludes.  But there are switches available

      \LabelsGlobal      \LabelsLocal

which change this.  To understand this, extend the above test 
by something like:

 \LabelsLocal

 \SetLabels
    (.5*.5) AAA\\
 \endSetLabels

 {
 \SetLabels
    (.5*.5) ZZZ\\
 \endSetLabels
   \AffixLabels{\FirstQuadrant}
 }

   \AffixLabels{\FirstQuadrant}

     There are however potential pitfalls.  Neither
labelfig.tex nor boxedeps.tex has been tested under extreme
conditions. Problems may occur if their procedures are
indiscriminately nested. For boxedeps.tex (not labelfig.tex)
there is a precise cause for worry, namely many of its
variables are "global", which means that TeX braces will not
provide the protection one might expect.

COMMAND SUMMARY FOR labelfig.tex

  Here [...] means optional (one or zero)
       [...]* means any number of such constructs

  \SetLabels
    [[<P>](<X><Sep><Y>) <label> \\]*
  \endSetLabels
  \ShowGrid  
  \AffixLabels{<the figure>}

   --- <P> is tack position, one of eleven or empty
              order irrelevant

                   \L\T      \T      \R\T

                   \L\E      \E      \R\E

                     \L               \R

                   \L\B      \B      \R\B

   --- (<X><Sep><Y>) insertion point;
  <Sep> is separator, = * by default;
  \SetXYSeparator{<Sep>} changes it.
   <X> and <Y> are real numbers

  --- <label> a label to attach 

  --- <the figure> the figure to label 

  \GlobalLabels (default)     
  \LocalLabels  setting for nested constructs.

 \Grids makes ALL grids appear; \HideGrid then makes just next disappear.
 \noGrids returns to default.  The commands are always global.

 \GridLineWidth{<dimension>} adjusts width of grid lines. Default is very
small, to give "hairline" effect. If your grid lines are missing try
setting \GridLineWidth{1pt}.

 \Edges#1 globally changes the edge size of all grids to the numerical 
value #1, which must be 1, 10, 100, or 1000.  The default is 1.

VERSION HISTORY.
 --- Jan 1993: \Edges#1 and [??] option after \SetLabels
 --- July 1992: \Grids, \noGrids, \HideGrid;
       Gridlines become hairlines; \GridLineWidth{<dimension>}.
 --- Oct 1991, Jan 1992: \SetXYSeparator{<Sep>},  \LabelsGlobal,
       \LabelsLocal.
 --- July 1991: first release

Address for bugs and other feedback:

        Raymond S\'eroul
        IREM and Lab. de Typographie Informatise
        Univ. Rene Descartes
        Strasbourg

    Tel 33-88-41-63-45
    Email:  A18645@FRCCSC21.BITNET

        Laurent Siebenmann
        Mathematique, Bat. 425,
        Univ de Paris-Sud,
        91405-Orsay,
        France

    Tel 33-1-6941-7949; 
    Email: lcs@topo.math.u-psud.fr

\def\scalefig#1{\epsfxsize #1\textwidth}
\def\defeq{\stackrel{\Delta}{=}}

\def\wt{\widetilde}

\def\diag{\text{diag}}
\def\ve{\text{vec}}

\def\blkToe{\text{blkToeplitz}}
\def\ovl{\overline}
\def\MSE{{\cal{M}}}

\newcommand {\Ebb}{{\mathbb{E}}}

\newcommand{\tr}{\mbox{${\mbox{tr}}$}}

\newtheorem{lemma}{Lemma}

\newtheorem{algorithm}{Algorithm}
\newtheorem{problem}{Problem}
\newtheorem{proposition}{Proposition}

\normalsize

\begin{document}
%
\title{ Filter-And-Forward Relay Design for MIMO-OFDM Systems}
%
%
%

\author{Donggun~Kim,~\IEEEmembership{Student Member,~IEEE,}
        Youngchul Sung$^\dagger$\thanks{$^\dagger$Corresponding author},
{\em Senior~Member, IEEE}, and
        Jihoon~Chung,~\IEEEmembership{Student Member,~IEEE,}
\thanks{The authors are with the Dept. of Electrical Engineering,  KAIST, Daejeon 305-701, South
Korea. E-mail:\{dg.kim@, ysung@ee., and j.chung@\}kaist.ac.kr.
This research was funded by the MSIP(Ministry of Science, ICT \& Future Planning), Korea in the ICT R\&D Program 2013.
}
}

%
%

\markboth{Submitted to IEEE Transactions on Communications, \today}%
{}
%



\maketitle

\begin{abstract}

In this paper, the filter-and-forward (FF) relay design for
multiple-input multiple-output (MIMO) orthogonal
frequency-division multiplexing (OFDM) systems is considered. Due
to the considered MIMO structure, the problem of joint design of
the linear MIMO transceiver at the source and the destination and
the FF relay at the relay is considered.  As the design criterion,
the minimization of weighted sum mean-square-error (MSE) is
considered first, and the joint design in this case is approached
based on alternating optimization that iterates  between optimal
design of the FF relay for a given set of MIMO precoder and
decoder and optimal design of the MIMO precoder and decoder for a
given FF relay filter. Next, the
joint design problem for rate maximization is considered based on
the obtained result regarding weighted sum MSE and the existing
result regarding the relationship between weighted MSE
minimization and rate maximization. Numerical results show the
effectiveness of the proposed FF relay design and significant
performance improvement by FF relays over widely-considered simple
AF relays for MIMO-ODFM systems.
\end{abstract}


%
\IEEEpeerreviewmaketitle

\section{Introduction}

Recently, the filter-and-forward (FF) relaying scheme has gained
an interest from the research communities as an alternative
relaying strategy due to its capability of performance improvement
over simple AF relays and still low complexity compared with other
relaying strategies such as decode-and-forward (DF) and
compress-and-forward (CF) schemes
\cite{ElGamal&Mohseni&Zahedi:06IT, DelCoso&Ibars:09WC,
Chen10:SP,Liang:11WCOM, SungKim:11arXiv,
KimSungLee:12SP,Dong:13VT}.  It is shown that the FF scheme can
outperform the AF scheme considerably. However, most of the works
regarding the FF relay scheme were conducted for single-carrier
systems
\cite{ElGamal&Mohseni&Zahedi:06IT,Chen10:SP,SungKim:11arXiv,
KimSungLee:12SP}. Recently, Kim {\it et al.} considered the FF
relay design for single-input and single-output (SISO) OFDM
systems \cite{DGKim:12APSIPA,Dong:13VT}, but their result based on
worst subcarrier signal-to-noise ratio (SNR) maximization or
direct rate maximization is not easily extended to the MIMO case
since SNR is not clearly defined for MIMO channels and furthermore
in the MIMO case the design of the MIMO precoder at the souce and
the MIMO decoder at the destination should be considered jointly
with the FF relay design. Thus, although there exists vast
literature regarding the relay design for MIMO-OFDM systems in the
case that the relay performs OFDM processing \footnote{In this case, each subcarrier channel is independent and we only need to consider a single flat MIMO channel.}
\cite{Hammerstr:06ConfCommun,Ng:07JSAC,Dong:10ICASSP,Dang:10WC,Fang&Hua&Koshy:06SAM,Simoens&Medina&Vidal&Coso:09SP},
not many results are available for the FF relay design for
MIMO-OFDM transmission, which is the current industry standard for
the physical layer of many commercial wireless communication
systems.

In this paper, we consider the FF relay design for MIMO-OFDM
systems. In the MIMO case, the FF relay should not be designed
alone without considering the MIMO precoder and decoder at the
source and the destination. Thus, we consider the problem of joint
design of the linear MIMO transceiver at the source and the
destination and the FF relay at the relay. As mentioned, in the
MIMO case, it is not easy to use SNR as the design metric as in
the SISO case \cite{Dong:13VT}. Thus, we approach the design
problem based on the tractable criterion of
 minimization of weighted sum MSE first
 and then consider the rate-maximizing design problem based on the
 equivalence relationship between rate maximization and weighted
 MSE minimization with a properly chosen weight matrix \cite{Sampath:01COM, Guo:05INF,
Palomar:06Inf,Cioffi:08WCOM,Luo:11SP}. We tackle the complicated
joint design problems by using alternating optimization, which
enables us to exploit the existing results for the MIMO precoder
and decoder design when all channel information is given. The
proposed alternating optimization is based on the iteration
between optimal design of the FF relay for a given set of MIMO
precoder and decoder and optimal design of the MIMO precoder and
decoder  for a given FF relay filter. While the linear MIMO
transceiver  design for a given FF relay filter can be addressed
by existing results e.g. \cite{Sampath:01COM}, the problem of
optimal design of the FF relay for a given MIMO transceiver is
newly formulated based on the block circulant matrix theorem and
reparameterization. It is shown that the FF relay design problem
for a given MIMO transceiver reduces to a quadratically
constrained quadratic program (QCQP) problem and a solution to
this QCQP problem is proposed based on conversion to a
semi-definite program (SDP). Numerical results show the
effectiveness of the proposed FF relay design and significant
performance improvement by FF relays over widely-considered simple
AF relays, and suggests that it is worth considering the FF
relaying scheme for MIMO-OFDM systems over the AF scheme with a
certain amount of complexity increase.

\subsection{Notation and Organization}
\vspace{-0.5em}

In this paper, we will make use of standard notational
conventions. Vectors and matrices are written in boldface with
matrices in capitals. All vectors are column vectors.  For a
matrix $\Xbf$, $\Xbf^*$, $\Xbf^T$, $\Xbf^H$, $\mbox{tr}(\Xbf)$,
and  $\Xbf(i,j)$ indicate the complex conjugate, transpose,
conjugate transpose, trace, and $(i,j)$-element of $\Xbf$,
respectively. $\Xbf \succeq 0$ and $\Xbf \succ 0$ mean that $\Xbf$
is positive semi-definite and that $\Xbf$ is strictly positive
definite, respectively. $\Ibf_n$ stands for the identity matrix of
size $n$ (the subscript is omitted
    when unnecessary),  ${\mathbf {I}}_{m \times n}$ denotes the first $m\times n$ submatrix of $\Ibf$, and  ${\mathbf {0}}_{m \times n}$ denotes a $m
\times n$ matrix  of all zero elements (the subscript is omitted
    when unnecessary).
The notation $\blkToe(\ovl{\Fbf},N)$ indicates a $N A \times
(N+L_f-1) B$ block Toeplitz matrix with $N $ row blocks  and $[
\ovl{\Fbf}, \bf{0}, \cdots, \bf{0} ] $ as its first row block,
where $\ovl{\Fbf} = [ \Fbf_0, \Fbf_1, \cdots, \Fbf_{L_f-1}] $ is a
row block composed of $A \times B$  matrices $\{\Fbf_k\}$;
$\diag(\Xbf_1, \Xbf_2, \cdots, \Xbf_n)$ means a (block) diagonal
matrix with diagonal entries $\Xbf_1, \Xbf_2, \cdots, \Xbf_n$. The
notation $\xbf\sim {\cal{CN}}(\mubf,\Sigmabf)$ means that $\xbf$
is complex circularly-symmetric Gaussian distributed with mean
vector $\mubf$ and covariance matrix $\Sigmabf$. $\Ebb\{\cdot\}$
denotes the expectation. $\iota=\sqrt{-1}$.

The remainder of this paper is organized as follows. The system
model is described in Section \ref{sec:systemmodel}. In Section
\ref{sec:ProblemFormulation}, the joint transceiver and FF relay
design problems for minimizing the weighted sum MSE and for
maximizing the data rate are formulated and solved by using convex
optimization theory and existing results. The performance of the
proposed design methods  is investigated in Section
\ref{sec:numericalresults}, followed by the conclusion in Section
\ref{sec:conclusion}.

\vspace{-0.8em}

\section{System Model}
\label{sec:systemmodel}

\begin{figure}[http]
\begin{psfrags}
\psfrag{s0}{\scriptsize$\sbf_0$} %
\psfrag{s1}{\scriptsize$\sbf_1$} %
\psfrag{sn}{\scriptsize$\sbf_{n}$} %
\psfrag{v0}{\scriptsize$\Vbf_0$} %
\psfrag{v1}{\scriptsize$\Vbf_1$} %
\psfrag{vn}{\scriptsize$\Vbf_{n}$} %
\psfrag{x0}{\scriptsize$\xbf_0$} %
\psfrag{x1}{\scriptsize$\xbf_1$} %
\psfrag{xn}{\scriptsize$\xbf_{n}$} %
\psfrag{a0}{\scriptsize$\Abf_0$} %
\psfrag{a1}{\scriptsize$\Abf_1$} %
\psfrag{ann}{\scriptsize$\Abf_{n}$} %
\psfrag{s00}{\scriptsize$\hat{\sbf}_0$} %
\psfrag{s11}{\scriptsize$\hat{\sbf}_1$} %
\psfrag{snn}{\scriptsize$\hat{\sbf}_{n}$} %
\psfrag{idft}{\scriptsize IDFT} %
\psfrag{dft}{\scriptsize DFT} %
\psfrag{ps}{\scriptsize P/S} %
\psfrag{an}{\scriptsize \&} %
\psfrag{cp}{\scriptsize CP} %
\psfrag{sp}{\scriptsize S/P} %
\psfrag{cpr}{\scriptsize CPR}%
\psfrag{ff}{\scriptsize FF relay} %
\psfrag{fir}{\scriptsize FIR filter} %
\psfrag{vd}{{\scriptsize$\vdots$}} %
\psfrag{Nt}{{\scriptsize$N_t$}} %
\psfrag{Mr}{{\scriptsize$M_r$}} %
\psfrag{Mt}{{\scriptsize$M_t$}} %
\psfrag{Nr}{{\scriptsize$N_r$}} %
\centerline{ \scalefig{0.95} \epsfbox{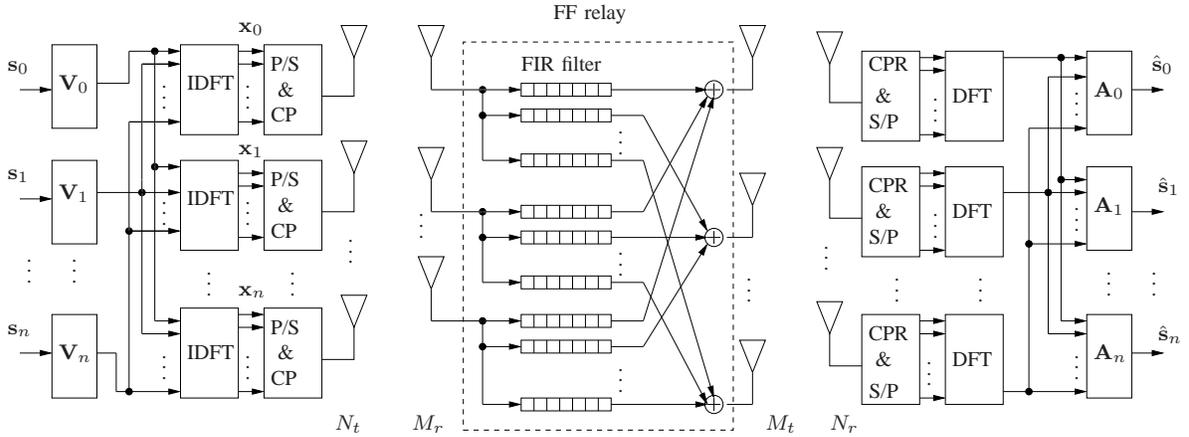} }
\captionsetup{justification=centering} \caption{System model}
\label{fig:system}
\end{psfrags}
\end{figure}

We consider a point-to-point MIMO-OFDM system with a
 relay, as shown in Fig.
\ref{fig:system}, where the source has $N_t$ transmit antennas,
the relay has  $M_r$ receive antennas and $M_t$ transmit antennas,
and the destination has $N_r$ receive antennas. The source and the
destination employ MIMO-OFDM modulation and demodulation with $N$
subcarriers, respectively, as in a conventional MIMO-OFDM system.
However, we assume that the relay is a full-duplex\footnote{In the
case of half-duplex, the problem can be formulated similarly.} FF
relay equipped with a bank of $M_t M_r$ finite impulse response
(FIR) filters with order $L_g$, i.e., the relay performs FIR
filtering on the incoming signals received at the $M_r$ receive
antennas at the chip rate\footnote{The FIR filtering is assumed to
be performed at the baseband. Thus, up and down converters are
necessary for FF operation and one common local oscillator (LO) at
the relay is sufficient.} of the OFDM modulation and transmits the
filtered signals instantaneously through the $M_t$ transmit
antennas to the destination without OFDM processing. Thus, the FF
relay can be regarded as an extension of an amplify-and-forward
(AF) relay and as an additional frequency-selective fading channel
between the source and the destination. We assume that there is no
direct link between the source and the destination and that the
source-to-relay (SR) and relay-to-destination (RD) channels are
multi-tap filters with finite impulse responses and their state
information  is known to the system.

The considered baseband system model is described in detail as
follows. At the source, a block of $N$ input data vectors of size
$\Gamma \times 1$, denoted as $\{\sbf_n = [s_n[1], s_n[2], \cdots,
s_n[\Gamma]]^T$, $n = 0, 1, \cdots, N-1$\}, is processed for one
OFDM symbol time. Here, $\sbf_n$ is the input data vector for the
 effective parallel flat MIMO channel at the $n$-th subcarrier provided by MIMO-OFDM
processing and $\Gamma \le \min(N_t, M_r, M_t, N_r)$ is the number
of data streams for the effective flat MIMO channel at each
subcarrier. We assume that each data symbol is a zero-mean
independent complex Gaussian random variable with unit variance,
i.e., $s_n[k] ~{\sim}~ {\cal{CN}}(0, 1)$ for $k= 1, 2, \cdots,
\Gamma$ and $n = 0, 1, \cdots, N-1$. Let the concatenated data
vector be denoted by $\sbf = [ \sbf_{N-1}^T, \sbf_{N-2}^T, \cdots,
\sbf_0^T ]^T$.  Although MIMO precoding can be applied to the
concatenated vector $\sbf$, such processing is complexity-wise
inefficient and thus we assume that MIMO precoding is applied to
the effective flat MIMO channel of each subcarrier separately, as
in most practical MIMO-OFDM systems, with a precoding matrix
$\Vbf_n$ for the $n$-th subcarrier MIMO channel. The MIMO precoded
$N$ symbols for each transmit antenna are collected and processed
by inverse discrete Fourier transform (IDFT). By concatenating all
IDFT symbols for all transmit antennas, we have the overall
time-domain signal vector $\xbf$, given by
\begin{equation}
\xbf  = ( \Wbf_{N} \otimes \Ibf_{N_t})\Vbf\sbf
\end{equation}
where
\begin{eqnarray}
\Vbf &=& \diag(\Vbf_{N-1}, \Vbf_{N-2}, \cdots, \Vbf_0)\\
\Wbf_{N} (k+1,l+1) &=&  \frac{1}{\sqrt{N}}  e^{\iota \frac{2\pi
kl}{N} }, ~~~k,l=0,1,\cdots, N-1,
\end{eqnarray}
and $\xbf$ is cyclic-prefix attached and transmitted. The cyclic
prefix attached signal vector $\xbf_{cp}$ can be expressed as
\begin{equation}  \label{eq:xbfcp}
\xbf_{cp} =  \underbrace{\left (\left [
\begin{array}{c}
\Ibf_{N} \\
\Ibf_{N_{cp}}~~{\bf{0}}  \\
\end{array}
\right ] \otimes \Ibf_{N_t} \right )}_{\defeq~\Tbf_{cp}} \xbf,
\end{equation}
where $N_{cp}$ is the cyclic prefix length, and ${\bf{0}}$ in
\eqref{eq:xbfcp} is an $N_{cp} \times (N - N_{cp})$ all-zero
matrix. We assume that the length of the overall FIR channel
between the source and the destination is not larger than that of
the OFDM cyclic prefix, i.e., $N_{cp} \ge L_f + L_r +L_g -3$,
where $L_f$, $L_r$, and $L_g$ denote the SR channel length, the
FIR filter  order at the relay, and the RD channel length,
respectively.

The transmitted signal
 $\xbf_{cp}$ passes through the SR channel, the relay FIR filter,
and the RD channel; is corrupted by white Gaussian noise; and  is
received at the destination. Then, the transmitted signal vector
at the relay and the received signal vector at the destination are
respectively given  by
\begin{equation} \label{eq:relaypower1}
\ybf_t = \Rbf\Fbf \xbf_{cp} + \Rbf\nbf_r~~ \text{and} ~~ \ybf_d =
\Gbf\Rbf\Fbf\xbf_{cp} + \Gbf\Rbf\nbf_r + \nbf_{d},
\end{equation}
where {\small
\begin{align}
&\ybf_d = \left [\ybf_{d,N-1}^T, \ybf_{d,N-2}^T, \cdots, \ybf_{d,0}^T \right]^T,\\
&\ybf_t = \left [\ybf_{t,N-1}^T, \ybf_{t,N-2}^T, \cdots, \ybf_{t,0}^T, \ybf_{t,-1}^T, \cdots ,\ybf_{t,-L_g+1}^T \right]^T,\\
&\xbf_{cp} = \left [ \xbf_{N-1}^T,  \xbf_{N-2}^T, \cdots,  \xbf_{0}^T, \xbf_{-1}^T,
\cdots,\xbf_{-L_g-L_r-L_f+3}^T \right]^T,  \\
&\nbf_r = \left [ \nbf_{r,N-1}^T,  \nbf_{r,N-2}^T, \cdots,  \nbf_{r,0}^T, \nbf_{r,-1}^T, \cdots, \nbf_{r,-L_g-L_r+2}^T \right]^T,\\
&\nbf_d = \left [ \nbf_{d,N-1}^T,  \nbf_{d,N-2}^T, \cdots,  \nbf_{d,0}^T \right]^T,\\
&\Gbf = \blkToe(\ovl{\Gbf},N),~\Rbf =
\blkToe(\ovl{\Rbf},N+L_g-1),~\Fbf = \blkToe(\ovl{\Fbf},N+L_g+L_r-2),\label{eq:firstrowblock22}\\
&\ovl{\Gbf} = [ \Gbf_0, \Gbf_1, \cdots, \Gbf_{L_g-1} ],~\ovl{\Rbf}
= [ \Rbf_0, \Rbf_1, \cdots, \Rbf_{L_r-1} ],~\ovl{\Fbf} = [ \Fbf_0,
\Fbf_1, \cdots, \Fbf_{L_f-1} ].\label{eq:firstrowblock}
\end{align}} \noindent
Here, $\ybf_{d,k}$ and $\nbf_{d,k}$ are $N_r \times 1$ vectors;
 $\ybf_{t,k}$ is a $M_t \times 1$ vector;  $\xbf_k$ is a $N_t \times 1$ vector; $\nbf_{r,k}$ is a $M_r \times 1$ vector;
$\Gbf_k$ is a $N_r \times M_t$ matrix; $\Rbf_k$ is a $M_t \times
M_r$ matrix; and $\Fbf_k$ is a $M_r \times N_t$ matrix.  The
entries of the noise vectors, $\nbf_{r,k}$ and $\nbf_{d,k}$, are
independently and identically distributed (i.i.d) Gaussian with
$\nbf_{r,k}[i] \stackrel{i.i.d.}{\sim} {\cal{CN}}(0, \sigma_r^2)$
and $\nbf_{d,k}[i] \stackrel{i.i.d.}{\sim} {\cal{CN}}(0,
\sigma_d^2)$. Then, the (cyclic-prefix portion removed) $N$-point
vector DFT of the received vector at the destination is given by
\begin{align}
& \ybf = ( \Wbf_{N}^H \otimes \Ibf_{N_r}) \Gbf \Rbf\Fbf\xbf_{cp} + ( \Wbf_{N}^H \otimes \Ibf_{N_r})\Gbf\Rbf\nbf_r + (\Wbf_{N}^H \otimes \Ibf_{N_r})\nbf_d,  \nonumber\\
& ~~=  ( \Wbf_{N}^H \otimes \Ibf_{N_r}) \Gbf \Rbf\Fbf\Tbf_{cp} ( \Wbf_{N} \otimes \Ibf_{N_t}) \Vbf\sbf + ( \Wbf_{N}^H \otimes \Ibf_{N_r})\Gbf\Rbf\nbf_r + (\Wbf_{N}^H \otimes \Ibf_{N_r})\nbf_d, \nonumber\\
&~~=  ( \Wbf_{N}^H \otimes \Ibf_{N_r}) \Hbf_c ( \Wbf_{N} \otimes \Ibf_{N_t}) \Vbf\sbf + ( \Wbf_{N}^H \otimes \Ibf_{N_r})\Gbf\Rbf\nbf_r + (\Wbf_{N}^H \otimes \Ibf_{N_r})\nbf_d, \label{eq:circulant} \\
&~~=  \Dbf \Vbf\sbf + ( \Wbf_{N}^H \otimes \Ibf_{N_r})\Gbf\Rbf\nbf_r + (\Wbf_{N}^H \otimes \Ibf_{N_r})\nbf_d, \label{eq:diagonal}
\end{align}
where $\ybf = [ \ybf_{N-1}^T, \ybf_{N-2}^T, \cdots, \ybf_{0}^T
]^T$, $\ybf_n$ is a $N_r \times 1$ received signal vector at the
$n$-th subcarrier, $\Wbf_{N}^H $ is the normalized DFT matrix of
size $N$, $\Hbf_c$ is a $N  N_r \times N  N_t$ block circulant
matrix generated from the block Toeplitz overall channel matrix
$\Gbf \Rbf\Fbf$ from the source to the destination, and $\Dbf = (
\Wbf_{N}^H \otimes \Ibf_{N_r}) \Hbf_c ( \Wbf_{N} \otimes
\Ibf_{N_t})$ is a block diagonal matrix generated by the block
circulant matrix theorem described in the next section. The $n$-th
subcarrier output of the $N$-point vector DFT is processed by a
linear receiver filter $\Ubf_n$ of size $\Gamma \times N_r$ to
yield an estimate of $\sbf_n$. The overall receiver processing for
all the subcarrier channels can be expressed as
\begin{align}
& \hat{\sbf} =  \Ubf\Dbf \Vbf\sbf + \Ubf( \Wbf_{N}^H \otimes
\Ibf_{N_r})\Gbf\Rbf\nbf_r + \Ubf(\Wbf_{N}^H \otimes
\Ibf_{N_r})\nbf_d,
\end{align}
where  $\Ubf = \diag(\Ubf_{N-1}, \Ubf_{N-2}, \cdots, \Ubf_0)$.

\subsection{Derivation of the subcarrier channel and mean square error}

To facilitate the optimization problem formulation in the next
section, we need to derive an explicit expression for the received
signal vector $\ybf_n$, $n = 0, 1, \cdots, N-1$, at the $n$-th
subcarrier.

\vspace{0.5em}
\begin{lemma} \label{lem:circulantM}
If $\Hbf_c$ is a block circulant matrix with $\Kbf = [ \Hbf_0,
\Hbf_1, \cdots, \Hbf_{N-1} ]$ as its first row block, then
it is block-diagonalizable as
\begin{equation*}
\Lambdabf_b = (\Wbf^H_N \otimes \Ibf_{N_r})~ \Hbf_c~ (\Wbf_N \otimes \Ibf_{N_t})
\end{equation*}
where $\Lambdabf_b $ is a block diagonal matrix defined as
\begin{equation*}
\Lambdabf_b  = \left [
\begin{array}{ccc}
\Kbf(\sqrt{N} \wbf_{N-1}^H \otimes \Ibf_{N_t})^T & & 0 \\
 & \ddots & \\
 0 & & \Kbf(\sqrt{N} \wbf_{0}^H \otimes \Ibf_{N_t})^T
\end{array}
\right ]
\end{equation*}
with $\sqrt{N}\wbf_k^H$ denoting the $-(k-N)$-th row of the DFT
matrix $\sqrt{N}\Wbf^H_N$, and
\begin{equation*}
\Kbf(\sqrt{N}\wbf_k^H \otimes \Ibf_{N_t})^T = \sum_{n=0}^{N-1}\Hbf_n~ e^{-\iota 2\pi \frac{n(N-k-1)}{N}}.
\end{equation*}
 \end{lemma}
\vspace{0.5em} \textit{Proof} : In \cite{Gray:06Toeplitzbook}, it
is shown that a circulant matrix can be diagonalized by a DFT
matrix. This can easily be extended to the block circulant case.
$\hfill{\square}$

By lemma 1, to derive the diagonal blocks of $\Dbf$ in
\eqref{eq:diagonal}, we only need to know the first row block of
$\Hbf_c$ in \eqref{eq:circulant}. Let the first row block of the
RD channel matrix $\Gbf$ be denoted by a $N_r \times M_t(N+L_g
-1)$ matrix $\wt{\Gbf} = [ \Gbf_0, \Gbf_1, \cdots, \Gbf_{L_g-1},
\bf{0}, \cdots, \bf{0}]$. Then, the first row block of the
effective channel filtering matrix $\Gbf\Rbf\Fbf$ is given by
$\wt{\Gbf} \Rbf \Fbf$. Note that the cyclic prefix adding and
removing operations make $\Gbf\Rbf\Fbf$ into the block circulant
matrix $\Hbf_c$ by truncating out the blocks of $\Gbf\Rbf\Fbf$
outside the first $N \times N$ blocks and by moving the lower
$(L_g + L_r +L_f -3) \times (L_g + L_r +L_f -3)$ blocks of the
truncated part to the lower left of the untruncated $N \times N$
block matrix, where each block is a $N_r \times N_t$ matrix.
Therefore, the first row block  $\wt{\Hbf}_c$ of $\Hbf_c$ is
simply the first $N$ blocks of $\wt{\Gbf} \Rbf \Fbf$, given by
\begin{equation}  \label{eq:tHderivationI}
\wt{\Hbf}_c = \wt{\Gbf} \Rbf \Fbf \Tbf ~~ \text{and}~~ {\Tbf}  = {\left [
\begin{array}{c}
\Ibf_{N N_t}\\
{\bf{0}}_{(L_f+L_r+L_g-3) N_t \times N N_t}
\end{array}
\right ] }
\end{equation}
where $\Tbf$ is a truncation matrix for truncating out the
remaining blocks of $\wt{\Gbf} \Rbf \Fbf$ except the first $N$
column blocks.  By using the first row block $\wt{\Hbf}_c$ and
Lemma 1, we  obtain the diagonal blocks of $\Dbf$ as
\begin{equation}  \label{eq:tHderivationII}
\Dbf = \diag(\wt{\Hbf}_c(\sqrt{N} \wbf_{N-1}^H \otimes
\Ibf_{N_t})^T, \wt{\Hbf}_c(\sqrt{N} \wbf_{N-2}^H \otimes
\Ibf_{N_t})^T, \cdots, \wt{\Hbf}_c(\sqrt{N} \wbf_{0}^H \otimes
\Ibf_{N_t})^T).
\end{equation}
Based on \eqref{eq:tHderivationI} and \eqref{eq:tHderivationII},
the received signal vector on the $n$-th subcarrier at the
destination is expressed as
\begin{align}
\ybf_n &= \sqrt{N}\wt{\Gbf}\Rbf\Fbf\Tbf\mathcal{W}_{t,n}^T \Vbf_n\sbf_n + \mathcal{W}_{r,n}\Gbf\Rbf\nbf_{r,n} + \mathcal{W}_{r,n}\nbf_{d,n}, \\
&= \hat{\ybf}_{n} + \zbf_n,
\end{align}
where $\mathcal{W}_{t,n} = \wbf_n^H\otimes \Ibf_{N_t}$,
$\mathcal{W}_{r,n} = \wbf_n^H\otimes \Ibf_{N_r}$, $~\hat{\ybf}_{n}
= \sqrt{N}\wt{\Gbf}\Rbf\Fbf\Tbf\mathcal{W}_{t,n}^T \Vbf_n\sbf_n$,
and $\zbf_n = \mathcal{W}_{r,n}\Gbf\Rbf\nbf_{r,n} +
\mathcal{W}_{r,n}\nbf_{d,n} $. This received signal vector
$\ybf_n$ is filtered by the receive filter $\Ubf_n$ and its output
is given by
\begin{equation} \label{eq:systemmodel1}
\hat{\sbf}_n =
\sqrt{N}\Ubf_n\wt{\Gbf}\Rbf\Fbf\Tbf\mathcal{W}_{t,n}^T
\Vbf_n\sbf_n + \Ubf_n\mathcal{W}_{r,n}\Gbf\Rbf\nbf_{r,n} +
\Ubf_n\mathcal{W}_{r,n}\nbf_{d,n}.
\end{equation}
Finally, the weighted MSE between $\sbf_n$ and $\hat{\sbf}_n$ is
given by
\begin{align}
\tr( \Thetabf_n \MSE_n )
&= \tr\left( \Thetabf_n\Ebb\left\{(\hat{\sbf}_n-\sbf_n)(\hat{\sbf}_n-\sbf_n)^H \right\} \right), \nonumber\\
&=  \tr\left( \Thetabf_n\Ebb\left\{(\Ubf_n\ybf_n-\sbf_n)(\Ubf_n\ybf_n-\sbf_n)^H \right\} \right), \nonumber\\
&= \tr \left( \Thetabf_n \left(\Ebb \{
\Ubf_n\ybf_n\ybf_n^H\Ubf_n^H \} -\Ebb\{\sbf_n\ybf_n^H\Ubf_n^H\}
-\Ebb\{\Ubf_n\ybf_n\sbf_n^H\}
+\Ebb\{\sbf_n\sbf_n^H\} \right) \right), \nonumber  \\
&= \tr \left( \Thetabf_n \Ebb \{\Ubf_n\hat{\ybf}_n\hat{\ybf}_n^H\Ubf_n^H \} \right )+ \tr \left( \Thetabf_n \Ebb \{ \Ubf_n\zbf_n\zbf_n^H\Ubf_n^H \} \right) -\tr \left( \Thetabf_n \Ebb\{\sbf_n\hat{\ybf}^H\Ubf_n^H\}\right) \nonumber \\
&~~~- \tr \left( \Thetabf_n
\Ebb\{\Ubf_n\hat{\ybf}_n\sbf_n^H\}\right) +\tr \left( \Thetabf_n
\Ebb\{\sbf_n\sbf_n^H\} \right),  \label{eq:MSE}
\end{align}
where
$\MSE_n\defeq\Ebb\left\{(\hat{\sbf}_n-\sbf_n)(\hat{\sbf}_n-\sbf_n)^H
\right\}$ is the MSE matrix at the $n$-th subcarrier and
$\Thetabf_n$ is a $\Gamma\times \Gamma$ diagonal positive definite
weight matrix.


\section{Problem Formulation and Proposed Design Method}  \label{sec:ProblemFormulation}


In this section, we consider optimal design of the FIR MIMO relay
filter $\{\Rbf_0,\Rbf_1,\cdots,\Rbf_{L_r-1}\}$ and the linear
precoders and decoders $\{\Vbf_n,\Ubf_n, n=0,1,\cdots,N-1\}$.
Among several optimality criteria, we first consider the
minimization of the weighted sum mean-square-error (MSE) for given
weight matrices, and then consider the rate maximization via the
weighted sum MSE minimization based on the fact that the rate
maximization for MIMO channels is equivalent to the weighted MSE
minimization with properly chosen weight matrices $\{\Thetabf_n\}$
\cite{Sampath:01COM}. (Here, the summation is across the
subcarrier channels.) The first problem is formally stated as
follows.

\vspace{0.3em}

\begin{problem} For given weight matrices $\{\Thetabf_n\}$,
SR channel $\Fbf$, RD channel $\Gbf$, FF relay filter order $L_r$,
maximum source transmit power $P_{s,max}$, and maximum relay
transmit power $P_{r,max}$, optimize the transmit filter
$\Vbf=\diag(\Vbf_0,\cdots,\Vbf_{N-1})$, the relay filter
$\ovl{\Rbf}$, and the receive filter
$\Ubf=\diag(\Ubf_0,\cdots,\Ubf_{N-1})$ in order to minimze the
weighted sum MSE:
\begin{eqnarray}
\underset{\Vbf, \ovl{\Rbf}, \Ubf} {\min} &&  \sum_{n = 0}^{N-1}
\tr(\Thetabf_n\MSE_n) ~~\mbox{s.t.} ~~ \tr(\Vbf\Vbf^H) \leq
P_{s,\max} ~~\text{and}~~ \tr(\ybf_t \ybf_t^H) \leq P_{r,\max}.
\end{eqnarray}
\end{problem}

\vspace{0.3em}

Note that Problem 1 is a complicated non-convex optimization
problem, which does not yield an easy solution. To circumvent the
difficulty in joint optimization, we approach the problem based on
alternating optimization.  That is, we first optimize the relay
filter for given transmit and receive filters under the power
constraints. Then, with the obtained relay filter we optimize the
transmit and receive filters.  Problem 1 is solved in this
alternating fashion until the iteration converges. A solution to
each step is provided in the following subsections.


\subsection{Relay Filter Optimization}


Whereas the linear precoder $\Vbf_n$ and decoder $\Ubf_n$ are
applied to each subcarrier channel separately, the relay filter
affects all the subcarrier channels simultaneously since the FF
relay does not perform OFDM processing. Here we consider the relay
filter optimization for given transmit and receive filters, and
the problem is formulated as follows.

\vspace{0.3cm} {\it Problem 1-1:} For given weight matrices
$\{\Thetabf_n\}$, SR channel $\Fbf$, RD channel $\Gbf$, FF relay
filter order $L_r$, transmit filter $\Vbf$,  receive filter
$\Ubf$, and maximum relay transmit power $P_{r,max}$, optimize the
relay filter $\ovl{\Rbf}$ in order to minimize the weighted sum
MSE:
\begin{equation} \label{eq:problem1d1}
\underset{\ovl{\Rbf} } {\min}   \sum_{n = 0}^{N-1}
\tr(\Thetabf_n\MSE_n) ~~\mbox{s.t.} ~~ \tr(\ybf_t \ybf_t^H) \leq
P_{r,\max}.
\end{equation}

\vspace{0.3em}

 To solve Problem 1-1, we first need to express each term in \eqref{eq:problem1d1} as a function of the design variable $\ovl{\Rbf}$.
Note that the relay block-Toeplitz filtering matrix $\Rbf$ is
redundant since the true design variable $\ovl{\Rbf}$ is embedded
in the block Toeplitz structure of $\Rbf$. (See
\eqref{eq:firstrowblock22}.) Hence, taking $\Rbf$ as the design
variable directly is inefficient and we need reparameterization of
the weighted MSE in terms of $\ovl{\Rbf}$. This is possible
through successive manipulation of the terms constructing the
weight MSE shown in \eqref{eq:MSE}. First, using similar techniques to those used in \cite{Dong:13VT}, we can express the
first term of \eqref{eq:MSE} in terms of $\ovl{\Rbf}$ as follows:
\begin{align}
&\tr(\Thetabf_n\Ebb\{\Ubf_n \hat{\ybf}_{n}\hat{\ybf}_{n}^H\Ubf_n^H\}) \nonumber\\
&= N \tr ( \Thetabf_n\Ebb\{ \Ubf_n\wt{\Gbf}\Rbf\Fbf\Tbf
\mathcal{W}_{t,n}^T
\Vbf_n\sbf_n\sbf_n^H\Vbf_n^H\mathcal{W}_{t,n}^*\Tbf^H\Fbf^H\Rbf^H\wt{\Gbf}^H\Ubf_n^H
     \} ), \nonumber\\
&\stackrel{(a)}{=}  N\mbox{tr} (\Vbf_n^H\mathcal{W}_{t,n}^*\Tbf^H\Fbf^H\Rbf^H
    \wt{\Gbf}^H\Ubf_n^H \Thetabf_n\Ubf_n\wt{\Gbf}
    \Rbf\Fbf\Tbf\mathcal{W}_{t,n}^T\Vbf_n\Ebb\left\{\sbf_n\sbf_n^H \right\} ),\nonumber\\
&= N\mbox{tr} (\Vbf_n^H\mathcal{W}_{t,n}^*\Tbf^H\Fbf^H\Rbf^H
    \wt{\Gbf}^H\Ubf_n^H \Thetabf_n\Ubf_n\wt{\Gbf}
    \Rbf\Fbf\Tbf\mathcal{W}_{t,n}^T\Vbf_n ),\nonumber\\
&= N ~ \mbox{tr} \big( \Thetabf_n^{1/2}\Ubf_n\wt{\Gbf} \Rbf
    \underbrace{\Fbf\Tbf\mathcal{W}_{t,n}^T\Vbf_n\Vbf_n^H\mathcal{W}_{t,n}^*\Tbf^H\Fbf^H}_{=:\Kbf_n}
    \Rbf^H\wt{\Gbf}^H\Ubf_n^H \Thetabf_n^{1/2}\big),\nonumber\\
&\stackrel{(b)}{=} N \left[\mbox{vec}(\Rbf^T\wt{\Gbf}^T\Ubf_n^T
\Thetabf^{1/2}_n)\right]^T
    \ovl{\Kbf}_n    \left[\mbox{vec}(\Rbf^T\wt{\Gbf}^T\Ubf_n^T\Thetabf^{1/2}_n)\right]^*, \nonumber\\
&\stackrel{(c)}{=}
     N \left[\mbox{vec}(\Rbf^T)\right]^T   (\Thetabf^{1/2}_n\Ubf_n\wt{\Gbf}\otimes \Ibf_Q)^T
    \ovl{\Kbf}_n    (\Thetabf^{1/2}_n\Ubf_n\wt{\Gbf}\otimes \Ibf_Q)^*  \left[\mbox{vec}(\Rbf^T)\right]^*,  \nonumber\\
&\stackrel{(d)}{=}
    N \rbf^T {\Ebf}_1 (\Thetabf^{1/2}_n\Ubf_n\wt{\Gbf}\otimes \Ibf_Q)^T
    \ovl{\Kbf}_n    (\Thetabf^{1/2}_n\Ubf_n\wt{\Gbf}\otimes \Ibf_Q)^* \Ebf_1^H\rbf^*, \nonumber\\
&= \rbf^H \Qbf_{1,n} \rbf,  \label{eq:MSE1stterm}
\end{align}
where
\begin{eqnarray}
\ovl{\Kbf}_n = { \Ibf_{\Gamma}\otimes\Kbf_n }; ~~ \Ibf_Q = \Ibf_{(N+L_r+L_g-2)M_r}; ~~\rbf = \ve(\ovl{\Rbf}^T); ~~~~~\nonumber\\
\Qbf_{1,n} =  N {\Ebf}_1^* (\Thetabf^{1/2}_n\Ubf_n\wt{\Gbf}\otimes
\Ibf_Q)^H
    \ovl{\Kbf}_n^*    (\Thetabf^{1/2}_n\Ubf_n\wt{\Gbf}\otimes \Ibf_Q) \Ebf_1^T;~~~~ \nonumber
\end{eqnarray}
and $\Ebf_1$ is defined in Appendix \ref{sec:appendix1}.
 Here, (a) holds due to $\tr(\Ubf\Bbf\Cbf) =
\tr(\Cbf\Ubf\Bbf)$; (b) holds due to $\tr(\Xbf \Kbf_n \Xbf^H) =
\ve(\Xbf^T)^T \ovl{\Kbf}_n \ve(\Xbf^T)^*$; (c) holds due to the
kronecker product identity,
$\mbox{vec}(\Ibf\Bbf\Cbf)=(\Cbf^T\otimes \Ibf)\mbox{vec}(\Bbf)$;
and (d) is obtained because $\Rbf = \blkToe(\ovl{\Rbf},N+L_g-1)$
and  $\mbox{vec}({\Rbf}^T) = \Ebf_1^T\rbf$.
 In a similar way, the remaining terms of \eqref{eq:MSE}  and the relay power constraint can also be represented as functions of the design variable
 $\rbf$. That is, the second term of \eqref{eq:MSE} can be
 rewritten as
\begin{align}
&\tr( \Thetabf_n\Ebb\left\{\Ubf_n\zbf_n\zbf_n^H\Ubf_n^H\right\} ) \nonumber \\
&=  \tr\left( \Thetabf_n\Ebb\left\{
\Ubf_n\mathcal{W}_{r,n}\Gbf\Rbf\nbf_{r,n}
   \nbf_{r,n}^H \Rbf^H\Gbf^H\mathcal{W}_{r,n}^H\Ubf_n^H  \right\}
    + \Thetabf_n\Ebb\left\{ \Ubf_n\mathcal{W}_{r,n}\nbf_{d,n} \nbf_{d,n}^H\mathcal{W}_{r,n}^H\Ubf_n^H \right\} \right), \nonumber \\
&=
\tr(\Rbf^H\Gbf^H\mathcal{W}_{r,n}^H\Ubf_n^H\Thetabf_n\Ubf_n\mathcal{W}_{r,n}\Gbf
    \Rbf \Ebb\left\{\nbf_{r,n}\nbf_{r,n}^H \right\} ) +\tr(\Thetabf_n\Ubf_n\mathcal{W}_{r,n}\Ebb\left\{\nbf_{d,n}^H\nbf_{d,n}\right\}\mathcal{W}_{r,n}^H\Ubf_n^H),
    \nonumber\\
&= \sigma_r^2 \mbox{tr} (\Rbf^H
    \underbrace{\Gbf^H\mathcal{W}_{r,n}^H\Ubf_n^H\Thetabf_n\Ubf_n\mathcal{W}_{r,n}\Gbf}_{=:\Mbf_n} \Rbf )
    +\sigma_d^2\mbox{tr}(\Thetabf_n\Ubf_n\mathcal{W}_{r,n}\mathcal{W}_{r,n}^H\Ubf_n^H), \nonumber\\
&= \sigma_r^2 \mbox{tr} (\Rbf^H    \Mbf_n  \Rbf )
    +\sigma_d^2\mbox{tr}(\Thetabf_n\Ubf_n(\wbf_n^H\otimes\Ibf_{N_r})(\wbf_n\otimes\Ibf_{N_r})\Ubf_n^H), \nonumber\\
&\stackrel{(a)}{=} \sigma_r^2 \ve(\Rbf)^H \ovl{\Mbf}_n
    \ve(\Rbf)
    +\sigma_d^2\mbox{tr}(\Thetabf_n\Ubf_n(\wbf_n^H\wbf_n\otimes\Ibf_{N_r})\Ubf_n^H), \nonumber \\
&\stackrel{(b)}{=} \sigma_r^2   \rbf^H\Ebf_2 \ovl{\Mbf}_n \Ebf_2^H \rbf
    +\sigma_d^2\mbox{tr}(\Thetabf_n\Ubf_n\Ubf_n^H),  \nonumber \\
&=   \rbf^H\Qbf_{2,n} \rbf
    + c_n,  \label{eq:MSEsecondterm}
\end{align}
where
\begin{equation*}
\ovl{\Mbf}_n= \Ibf_{(N+L_g+L_r-2)M_r} \otimes \Mbf_n,~~ \Qbf_{2,n}
= \sigma_r^2 \Ebf_2 \ovl{\Mbf}_n \Ebf_2^H, ~~ c_n =
\sigma_d^2\mbox{tr}(\Thetabf_n\Ubf_n\Ubf_n^H),
\end{equation*}
 and $\Ebf_2$ is defined  in Appendix \ref{sec:appendix1}. Here, (a) follows from the kronecker product identity
$(\Ubf\Bbf \otimes \Cbf\Dbf) = (\Ubf \otimes \Cbf)(\Bbf \otimes
\Dbf)$, and (b) is obtained due to $\ve(\Rbf)^H = \rbf^H\Ebf_2$.
The third term of \eqref{eq:MSE} can be
 rewritten as
\begin{align}
\tr \left( \Thetabf_n \Ebb\{\sbf_n\hat{\ybf}_n^H\Ubf_n^H\}\right)
&= \sqrt{N}\mbox{tr} \left(\Thetabf_n  \Ebb\left\{
    \sbf_n\sbf_n^H \right\}\Vbf_n^H\mathcal{W}_{t,n}^*\Tbf^H \Fbf^H \Rbf^H \wt{\Gbf}^H \Ubf_n^H  \right), \nonumber \\
&= \sqrt{N}\mbox{tr} \left(\Thetabf_n \Vbf_n^H\mathcal{W}_{t,n}^*\Tbf^H \Fbf^H \Rbf^H \wt{\Gbf}^H \Ubf_n^H  \right), \nonumber \\
&= \sqrt{N}\mbox{tr} \left(\Rbf^H \wt{\Gbf}^H \Ubf_n^H\Thetabf_n \Vbf_n^H\mathcal{W}_{t,n}^*\Tbf^H \Fbf^H   \right), \nonumber \\
&= \sqrt{N} \ve(\Rbf)^H \ve(\wt{\Gbf}^H \Ubf_n^H\Thetabf_n \Vbf_n^H\mathcal{W}_{t,n}^*\Tbf^H \Fbf^H), \nonumber  \\
&= \sqrt{N} \rbf^H \Ebf_2 \ve(\wt{\Gbf}^H \Ubf_n^H\Thetabf_n \Vbf_n^H\mathcal{W}_{t,n}^*\Tbf^H \Fbf^H), \nonumber  \\
&= \rbf^H \qbf_n,  \label{eq:MSEthirdterm}
\end{align}
where $\qbf_n = \sqrt{N}\Ebf_2 \ve(\wt{\Gbf}^H \Ubf_n^H\Thetabf_n
\Vbf_n^H\mathcal{W}_{t,n}^*\Tbf^H \Fbf^H)$. Finally, the relay
transmit power can be rewritten as
\begin{align}
&\Ebb\{\tr(\ybf_{t}\ybf_{t}^H )\} \nonumber \\
&= \tr\left( \Rbf \Fbf\Tbf_{cp}( \Wbf_{N} \otimes \Ibf_{N_t}) \Vbf\Ebb\{\sbf \sbf^H \} \Vbf^H ( \Wbf_{N}^H \otimes \Ibf_{N_t}) {\Tbf_{cp}}^H \Fbf^H \Rbf^H \right) +\tr \left( \Rbf \Ebb\{\nbf_{r}\nbf_r^H\}\Rbf^H \right), \nonumber \\
&= \tr\left( \Rbf \Fbf\Tbf_{cp}( \Wbf_{N} \otimes \Ibf_{N_t}) \Vbf \Vbf^H ( \Wbf_{N}^H \otimes \Ibf_{N_t}) {\Tbf_{cp}}^H \Fbf^H \Rbf^H \right) + \sigma^2_r\tr \left( \Rbf \Rbf^H \right), \nonumber \nonumber\\
&= \mbox{tr} \left( \Rbf \underbrace{(   \Fbf\Tbf_{cp}( \Wbf_{N} \otimes \Ibf_{N_t}) \Vbf \Vbf^H ( \Wbf_{N}^H \otimes \Ibf_{N_t}) {\Tbf_{cp}}^H \Fbf^H+\sigma^2_r\Ibf)}_{\Pibf} \Rbf^H  \right), \nonumber\\
&= \mbox{vec}(\Rbf^T)^T \ovl{\Pibf} \mbox{vec}(\Rbf^T)^*, \nonumber\\
&= \rbf^H\wt{\Pibf} \rbf, \label{eq:relaypower2}
\end{align}
where $\ovl{\Pibf} =   \Ibf_{(N+L_g -1)M_t} \otimes \Pibf$ and $\wt{\Pibf} = \Ebf_1^*  \ovl{\Pibf}^* \Ebf_1^T$.

Based on \eqref{eq:MSE1stterm}, \eqref{eq:MSEsecondterm},
\eqref{eq:MSEthirdterm}, and   \eqref{eq:relaypower2}, the
weighted MSE for the $n$-th subcarrier channel is expressed as
\begin{equation} \label{eq:MSE2}
\tr( \Thetabf_n \MSE_n ) = \rbf^H  \Qbf_n \rbf  - \rbf^H \qbf_n  - \qbf_n^H\rbf + z_n
\end{equation}
where $ \Qbf_n = \Qbf_{1,n} + \Qbf_{2,n}$ and $z_n = c_n +
\tr(\Thetabf_n)$, and  Problem 1-1 is reformulated as
\begin{eqnarray} \label{eq:QCQP}
\underset{\rbf}{\min} && \rbf^H \Qbf \rbf - \rbf^H \qbf  - \qbf^H\rbf + z   \nonumber \\
\mbox{s.t.} && \rbf^H\wt{\Pibf} \rbf \leq P_{r,\max}.
\end{eqnarray}
where $ \Qbf = \sum_{n=1}^N \Qbf_n $, $ \qbf = \sum_{n=1}^N \qbf_n $, and $ z = \sum_{n=1}^N z_n $.

 The key point of the derivation of \eqref{eq:QCQP} is that Problem 1-1 reduces to a {\em quadratically constrained quadratic programming
(QCQP) problem} with a constraint. It is  known that QCQP is
NP-hard in general. However, QCQP has been well studied in the
case that the number of constraints is small. Using the results of
\cite{Huang:Math07} and \cite{Wenbao:Math07}, we obtain an optimal
solution to Problem 1-1 as follows.  Let  $\ovl{\rbf}/{t} = \rbf$,
where $t \in {\cal{C}}$, and $\wt{\rbf} = [ \ovl{\rbf}^T, t ]^T
\in {\cal{C}}^{(M_t L_r M_r + 1) \times 1}$. Then, we rewrite
\eqref{eq:QCQP} equivalently as
\begin{eqnarray}
\underset{\wt{\rbf}}{\min} && \wt{\rbf}^H \Bbf_1 \wt{\rbf}  \nonumber \\
\mbox{s.t.} && \wt{\rbf}^H \Bbf_2 \wt{\rbf} \le 0
\end{eqnarray}
where
\begin{equation*}
\Bbf_1 =\left[
\begin{array}{cc}
\Qbf& -\qbf\\
-\qbf^H & z
\end{array}
 \right]
~~ \text{and}~~
\Bbf_2 =\left[
\begin{array}{cc}
\wt{\Pibf}& \bf{0}\\
\bf{0} & - P_{r,max}
\end{array}
 \right].
\end{equation*}
By defining  ${\cal{R}} := \wt{\rbf}\wt{\rbf}^H$ and removing the
rank-one constraint $\text{rank}( {\cal{R}} ) =1 $, we obtain the
following convex optimization problem:
\begin{eqnarray} \label{problem:SDP}
\underset{{\cal{R}}}{\min} && \tr(\Bbf_1 {\cal{R}})  \nonumber \\
\mbox{s.t.} && \tr(\Bbf_2 {\cal{R}}) \le 0
\end{eqnarray}
which is a semi-definite program (SDP) and can be solved
efficiently by using the standard interior point method for convex
optimization \cite{Boyd:04cvxbook, Helmberg:02SDP,
Sturm:99OptSeduMi, Boyd:CVX}. With an additional constraint
$\text{rank}( {\cal{R}} ) =1 $, the problem \eqref{problem:SDP} is
equivalent to Problem 1-1. That is, if the optimal solution of
\eqref{problem:SDP} has rank one, then it is also the optimal
solution of Problem 1-1. However, there is no guarantee that an
algorithm for solving the problem \eqref{problem:SDP} yields a
rank-one solution. In such a case, a rank-one solution from
$\cal{R}$ can always be obtained by using the rank-one
decomposition procedure \cite{Wenbao:Math07}.

\subsection{Transmit and receive filter optimization}

Now consider the joint design of the transmit and receive filters
$\{(\Vbf_n,\Ubf_n), n=0,1,\cdots,N-1\}$ for a given relay FIR
filter. Note that when the transmit power $P_{n,max} ~(\ge
\mbox{tr}(\Vbf_n\Vbf_n^H))$ for each $n$ and the relay filter are
 given,  the problem
simply reduces to $N$ independent problems of designing the
transmit filter $\Vbf_n$ and the receive filter $\Ubf_n$ for the
$n$-th subcarrier MIMO channel for $n = 0, \cdots, N-1$, as in typical MIMO-OFDM systems. This is
because we get an independent MIMO channel per subcarrier owing to
MIMO-OFDM processing. However, we have an additional freedom to
distribute the total source transmit power $P_{s,\max}$ to $N$
subcarriers such that $P_{s,\max}=\sum_{n=0}^{N-1} P_{n,\max}$,
and should take this overall power allocation into consideration.  So, we solve this problem by
separating the power allocation problem out and applying the
existing result \cite{Sampath:01COM} to this problem.  First,
consider the transmit and receive filter design problem when the
transmit power $P_{n,max}$ for each $n$ and the relay filter are
 given:

\vspace{0.3cm} {\it Problem 1-2:} For given weight matrices
$\{\Thetabf_n\}$,  maximum per-subcarrier transmit power
$P_{n,max}$ for $n= 0, 1, \cdots, N-1$, SR channel $\Fbf$, RD
channel $\Gbf$, relay filtering matrix $\Rbf$, jointly optimize
$(\Vbf_n,\Ubf_n)$ in order to minimize the weighted MSE at the
$n$-th subcarrier MIMO channel:
\begin{eqnarray}
\underset{ \Vbf_n, \Ubf_n } {\min} &&  \tr(\Thetabf_n\MSE_n)
~~\mbox{s.t.} ~~ \tr(\Vbf_n \Vbf_n^H) \leq P_{n,\max} , ~~
\text{for}~ n = 0, 1, \cdots, N-1.
\end{eqnarray}

Problem 1-2 has already been solved and  the optimal transceiver
structure for Problem 1-2 is available in \cite{Sampath:01COM}
and \cite{Sampath:Conf99}. It is shown in \cite{Sampath:01COM} that
the optimal transmit filter and receive filter diagonalize the
MIMO channel into eigen-subchannels for any weight matrix.  Lemma
1 and Theorem 1 of \cite{Sampath:01COM} provide  the optimal
transmit filter $\Vbf_n$ and receive filter $\Ubf_n$, and the
solution can be expressed as $\Vbf_n = \wt{\Vbf}_n\wt{\Pbf}_n$,
where $\wt{\Vbf}_n^H\wt{\Vbf}_n=\Ibf_{\Gamma}$ and $\wt{\Pbf}_n$
is a diagonal matrix with nonnegative entries s.t.
$\mbox{tr}(\wt{\Pbf}_n^2)=P_{n,\max}$ determining the transmit
power of each of $\Gamma$ data streams of the $n$-th subcarrier
MIMO channel. (Please refer to \cite{Sampath:01COM}.)

Note that the solution to Problem 1-2 only optimizes the power
allocation within multiple data streams for each subcarrier when
the transmit power is allocated to each subcarrier.  Now, consider
the problem of total source power allocation $P_{s,\max}$ to
subcarrier channels. Here, we exploit the {\em diagonalizing}
property \cite{Sampath:01COM} of the solution to Problem 1-2,   take the direction information only for the transmit filter from the solution to Problem 1-2, and apply alternating
optimization. That is, when the relay filtering matrix $\Rbf$ from
Problem 1-1 and the normalized transmit filters $\{\wt{\Vbf}_n\}$
and the receive filters $\{\Ubf_n\}$ from Problem 1-2 are given,
each subcarrier MIMO channel is diagonalized into
eigen-subchannels. Thus, the effective parallel MIMO channel
\eqref{eq:systemmodel1} for the $n$-th subcarrier is rewritten as
    \begin{align}
\hat{\sbf}_n &= \sqrt{N}\Ubf_n\wt{\Gbf}\Rbf\Fbf\Tbf\mathcal{W}_{t,n}^T \Vbf_n\sbf_n + \Ubf_n\mathcal{W}_{r,n}\Gbf\Rbf\nbf_{r,n} + \Ubf_n\mathcal{W}_{r,n}\nbf_{d,n} \nonumber \\
   &= \sqrt{N}\Ubf_n\wt{\Gbf}\Rbf\Fbf\Tbf\mathcal{W}_{t,n}^T \wt{\Vbf}_n \wt{\Pbf}_n\sbf_n + \Ubf_n\mathcal{W}_{r,n}\Gbf\Rbf\nbf_{r,n} + \Ubf_n\mathcal{W}_{r,n}\nbf_{d,n}, ~~~ \label{eq:subcarrierChannel}  \\
   &= \Dbf_n \wt{\Pbf}_n\sbf_n + \Ubf_n\mathcal{W}_{r,n}\Gbf\Rbf\nbf_{r,n} + \Ubf_n\mathcal{W}_{r,n}\nbf_{d,n}
\end{align}
where  $\Dbf_n = \diag(d_n[1], d_n[2], \cdots, d_n[\Gamma]$) is
obtained from the optimal transceiver $(\tilde{\Vbf}_n,\Ubf_n)$ of
Problem 1-2 with each $d_n[k]$ being a non-negative value
\cite{Sampath:01COM}, and $\wt{\Pbf}_n = \diag( p_n[1], p_n[2],
\cdots, p_n[\Gamma]$).  Therefore, we obtain $N \Gamma$ parallel
eigen-subchannels for the overall MIMO-OFDM system as
    \begin{equation}
    \hat{s}_n[k] = d_n[k] p_n[k] s_n[k] + n_n[k],~~\text{for}~n = 0, 1, \cdots, N-1 ~\text{and}~ k = 1, 2, \cdots,
    \Gamma,
    \end{equation}
    where  $n_n[k] = \Ubf_{n,k}^H \mathcal{W}_{r,n}(\Gbf\Rbf\nbf_{r,n} +\nbf_{d,n}) $ and $\Ubf_{n,k}^H$ is the $k$-th row of $\Ubf_n$.
The total power $P_{s,\max}$ should now be  optimally allocated to
these $N\Gamma$ parallel channels to minimize the weighted sum
MSE, where the weighted sum MSE of $N\Gamma$ parallel eigen-subchannels
is derived as
\begin{equation}
\sum_{n=0}^{N-1}\sum_{k=1}^B \theta_{nk}\Ebb \{ |\hat{s}_n[k] - s_n[k] |^2 \} =  \sum_{n=0}^{N-1}\sum_{k=1}^\Gamma \theta_{nk}  (d_n[k]^2 p_n[k]^2 - 2 d_n[k]  p_n[k] + c_n[k] )\\
\end{equation}
    where $c_n[k] = \sigma_r^2\Ubf_{n,k}^H \mathcal{W}_{r,n}\Gbf\Rbf\Rbf^H\Gbf^H \mathcal{W}_{r,n}^H \Ubf_{n,k} + \sigma_d^2 \Ubf_{n,k}^H
    \Ubf_{n,k}+1$, and $\theta_{nk}$ is properly derived from
    $\Thetabf_n$. Thus, the problem of overall source power allocation to
minimize the weight sum MSE subject to the source power constraint
is stated as follows.

\vspace{0.3cm} {\it Problem 1-3:} For given any weight matrices
$\{\Thetabf_n\}$, SR channel $\Fbf$, RD channel $\Gbf$, relay
filtering matrix $\Rbf$, maximum source power $P_{s,max} =
\sum_{n=0}^{N-1} P_{n,max}$,
 normalized transmit filters $\{\wt{\Vbf}_n\}$, and receive filters
$\{\Ubf_n\}$,
\begin{eqnarray} \label{problem:sourcePA}
\underset{p_n[k]} {\min} &&   \sum_{n=0}^{N-1}\sum_{k=1}^\Gamma
\theta_{nk} (d_n[k]^2 p_n[k]^2 - 2 d_n[k]  p_n[k] + c_n[k] )
~~\mbox{s.t.} ~~ \sum_{n=0}^{N-1}\sum_{k=1}^\Gamma p_n[k]^2 =
P_{s,max}.
\end{eqnarray}
Note that Problem 1-3 is a convex optimization problem with
respect to $p_n[k]$. The optimal solution to Problem 1-3 is given
in the following proposition:

\vspace{0.3em}
\begin{proposition}
The optimal solution to Problem 1-3 is given by
\begin{equation}
p_n[k] = \left (\frac{\theta_{nk}d_n[k]}{\theta_{nk}d_n[k]^2 +
\mu} \right)_+~~\text{s.t.}~~ \sum_{n=0}^{N-1}\sum_{k=1}^\Gamma
\left  (\frac{\theta_{nk}d_n[k]}{\theta_{nk}d_n[k]^2 + \mu} \right
)^2 = P_{s,max}.
\end{equation}
\end{proposition}
{\it{Proof}} : See  Appendix \ref{sec:appendix2}

\vspace{0.3em}

The solution in Proposition 1 allocates power inverse-proportionally
to the power of the effective channel $d_n[k]$ in most cases
similarly to the method in \cite{Sampath:Conf99}.

Now summarizing the results, we propose our method to design the
linear transceiver at the source and the destination and the FF
relay filter jointly to minimize the weighted sum MSE, based on
alternating optimization solving Problem 1-1, Problem 1-2, and
Problem 1-3 iteratively.

\begin{algorithm} Given parameters: $\{\Thetabf_n\}$, $\Fbf$, $\Gbf$, $L_r$, $P_{s,max}$, and $P_{r,max}$ \\
Step 1: Initialize $\{\wt{\Pbf}_n\}$, $\{\wt{\Vbf}_n\}$, and
$\{\Ubf_n\}$ for $n=0, 1, \cdots, N-1$. For example, $p_n[k] = \frac{P_{s,max}}{N\Gamma}$, $\wt{\Vbf}_n = \Ibf_{N_t \times \Gamma}$, and $\Ubf_n = \Ibf_{\Gamma\times N_r}$.\\
Step 2: Solve Problem 1-1 and obtain  $\Rbf$. \\
Step 3: Solve Problem 1-2 and obtain  $\{\wt{\Vbf}_n,\Ubf_n\}$. \\
Step 4: Solve Problem 1-3 and obtain  $\{\wt{\Pbf}_n\}$. \\
Step 5: Go to Step 2 and repeat until the change in the weighted sum MSE falls within a given tolerance.\\
\end{algorithm}
\vspace{-0.7cm} \noindent The weighted sum MSE is a function of
$\Rbf$ and $\{\wt{\Vbf}_n, \Ubf_n, \wt{\Pbf}_n\}$ denoted by
${\cal{M}}(\Rbf, \wt{\Vbf}_n, \Ubf_n, \wt{\Pbf}_n)$.  Let
$\Xbf^{(i)}$ denotes the solution at the $(i)$-th step. Then, it
is easy to see that ${\cal{M}}(\Rbf^{(0)}, \wt{\Vbf}_n^{(0)},
\Ubf_n^{(0)}, \wt{\Pbf}_n^{(0)})$  $\ge {\cal{M}}(\Rbf^{(1)},
\wt{\Vbf}_n^{(0)}, \Ubf_n^{(0)}, \wt{\Pbf}_n^{(0)}) \ge
{\cal{M}}(\Rbf^{(1)}, \wt{\Vbf}_n^{(2)}, \Ubf_n^{(2)},
\wt{\Pbf}_n^{(0)}) \ge {\cal{M}}(\Rbf^{(1)}, \wt{\Vbf}_n^{(2)},
\Ubf_n^{(2)}, \wt{\Pbf}_n^{(3)}) \ge  \cdots \ge 0$ because the
optimal solution is obtained at each step and the possible
solution set of the current step includes the solution of the
previous step. In this way, the proposed algorithm converges by
the monotone convergence theorem although it yields a suboptimal
solution and the initialization of the algorithm affects its
performance.

\subsection{Rate maximization}

Now we consider the problem of rate maximization.   In general,
the rate maximization problem is not equivalent to the MSE
minimization problem. However, they are closely related to each
other. The relationship has been studied in \cite{Sampath:01COM,
Guo:05INF, Palomar:06Inf}. By using the relationship, the rate
maximization problem for MIMO broadcast channels and MIMO
interference-broadcast channels has recently been considered in
\cite{Cioffi:08WCOM} and \cite{Luo:11SP}. In the case of the joint
design of the FF relay at the relay and the linear transceiver at
the source and the destination, the result regarding the weighted
sum MSE minimization in the previous subsection can be modified
and used to maximize the sum rate based on the existing
relationship between the weighed MSE and the rate. It was shown in
\cite{Sampath:01COM} that the rate maximization for the $n$-th
subcarrier MIMO channel \eqref{eq:subcarrierChannel} is equivalent
to the weighted MSE minimization when the weight matrix
$\Thetabf_n$ is set as a diagonal matrix composed of the
eigenvalues of $\Hbf^H \Sigmabf_{n}^{-1}\Hbf$, where $\Hbf=
\sqrt{N}\wt{\Gbf}\Rbf\Fbf\Tbf\mathcal{W}_{t,n}^T $ is the effective MIMO
channel matrix and $\Sigmabf_{n}$ is the effective noise
covariance matrix of the $n$-th subcarrier MIMO channel
\eqref{eq:subcarrierChannel}. (See Lemma 3 of
\cite{Sampath:01COM}.)  Exploiting this result, we propose our
algorithm to design the linear transceiver and the relay filter to
maximize the sum rate below.

\vspace{0.3em}

\begin{algorithm} Given parameters: $\Fbf$, $\Gbf$, $L_r$, $P_{s,max}$, and $P_{r,max}$ \\
Step 1: Initialize $\{\Thetabf_n\}$, $\{\wt{\Pbf}_n\}$,
$\{\wt{\Vbf}_n\}$, and
$\{\Ubf_n\}$ for $n=0, 1, \cdots, N-1$. For example, $\Thetabf_n = \Ibf$, $p_n[k] = \frac{P_{s,max}}{N\Gamma}$, $\wt{\Vbf}_n = \Ibf_{N_t \times \Gamma}$, and $\Ubf_n = \Ibf_{\Gamma\times N_r}$.\\
Step 2: Solve Problem 1-1 and obtain  $\Rbf$. \\
Step 3: \footnote{When $\Rbf$ is given, all the parallel subcarrier MIMO channels are determined and a solution  $\{\wt{\Vbf}_n,\Ubf_n,\Thetabf_n\}$ is given by Lemma 1 and Theorem 1 of \cite{Sampath:01COM}.}Solve Problem 1-2 and obtain  $\{\wt{\Vbf}_n,\Ubf_n,\Thetabf_n\}$. \\
Step 4: Compute  $\{\wt{\Pbf}_n\}$ for the $N\Gamma$ parallel scalar channels obtained from Step 3 by water-filling. \\
Step 5: Go to Step 2 and repeat until the change in the weighted sum MSE falls within a given tolerance.\\
\end{algorithm}

\vspace{0.3em}

\noindent Note that the weight matrices $\{\Thetabf_n\}$ in
Algorithm 2 are updated in each iteration so that the weighted MSE
minimization is equivalent to the rate maximization for an updated
relay filter, whereas the weight matrices are fixed over
iterations in Algorithm 1.

Now consider the complexity of the proposed algorithms.  Note that
solving Problem 1-2 involves $N$ separate small MIMO systems of
size $N_r \times N_t$, and the solution to Problem 1-3 (Algorithm
1) and the water-filling power allocation solution (Algorithm 2)
are explicitly given. Thus, the main complexity of the proposed
algorithms lies in solving Problem 1-1 that requires solving an
SDP problem of size $M_tM_rL_g$. Due to the existence of fast
approximate algorithms for solving SDP problems
\cite{Arora:Conf05, Arora:12TOC}, the proposed algorithm is
implementable if the number of iterations for convergence is not
so large, which will be seen in Fig. \ref{fig:Converge}. For other
practical issues such as channel estimation and self-interference
caused by full-duplex operation, please see \cite{Dong:13VT}.


\section{Numerical results}    \label{sec:numericalresults}


 In this section, we provide some numerical results to evaluate the
 performance of the proposed FF relay design in Section \ref{sec:ProblemFormulation}. Throughout the simulation, we fixed the number  of OFDM subcarriers
 as $N=16$ with a minimal cyclic prefix covering the overall
FIR channel length in each simulation case. In all cases, each
channel tap coefficient of the SR and RD channel matrices,
$\Fbf_k$ and $\Gbf_k$, was  generated  i.i.d according to  a
Rayleigh distribution, i.e., $\Fbf_k(i,j) \stackrel{i.i.d.}{\sim}
{\cal{CN}}(0, \sigma_f^2)$ and $\Gbf_k(i,j)
\stackrel{i.i.d.}{\sim} {\cal{CN}}(0, \sigma_g^2)$, where
$\sigma_f = \sigma_g = 1$.  The SR channel length and the RD
channel length were set as $L_f = L_g = 3$, and  $N_t = M_r = M_t
= N_r = 2$.  The relay and the destination had the same noise
power $\sigma_r^2 = \sigma_d^2 = 1$, and the source transmit power
was 20 dB higher than the noise power, i.e., {$P_{s,max}=100$.
(From here on, all dB power values are relative to
$\sigma_r^2=\sigma_d^2=1$.)}

\begin{figure}[http] \centerline{
    \begin{psfrags}
      \scalefig{0.6}\epsfbox{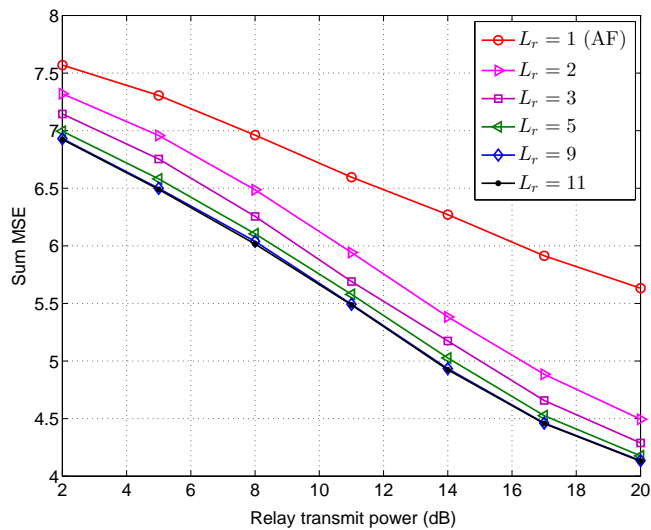}
    \end{psfrags}
 }
\captionsetup{justification=centering} \caption{Sum MSE versus FF relay transmit
power.} \label{fig:Sum_MSE}
\end{figure}

\begin{figure}[http] \centerline{
    \begin{psfrags}
      \scalefig{0.6}\epsfbox{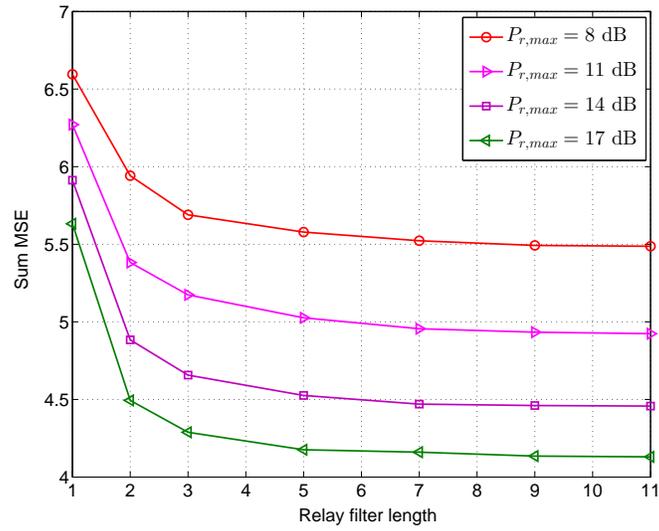}
    \end{psfrags}
} \captionsetup{justification=centering} \caption{Sum MSE versus relay filter
length.} \label{fig:Sum_MSE2}
\end{figure}

We first evaluated the MSE performance of the proposed FF relay
design
 method, Algorithm 1, to minimize the sum MSE subject to a source
power constraint and a relay
 power constraint.  Figs. \ref{fig:Sum_MSE} and \ref{fig:Sum_MSE2} show the resulting sum MSE over all subcarriers.
For the curves in the figures, 200 channels were randomly realized
with $L_f=L_g=3$ and each plotted value is the average over the
200 channel realizations.  As expected, it is seen in Figs.
\ref{fig:Sum_MSE} and \ref{fig:Sum_MSE2} that the performance of
the FF relay improves as the FF relay filter length increases, and the FF relay significantly outperforms the
simple AF relay ($L_r = 1$). It
is also seen that    most of the gain is achieved by
only a few filter taps for the FF relay.

\begin{figure}[http] \centerline{
    \begin{psfrags}
      \scalefig{0.6}\epsfbox{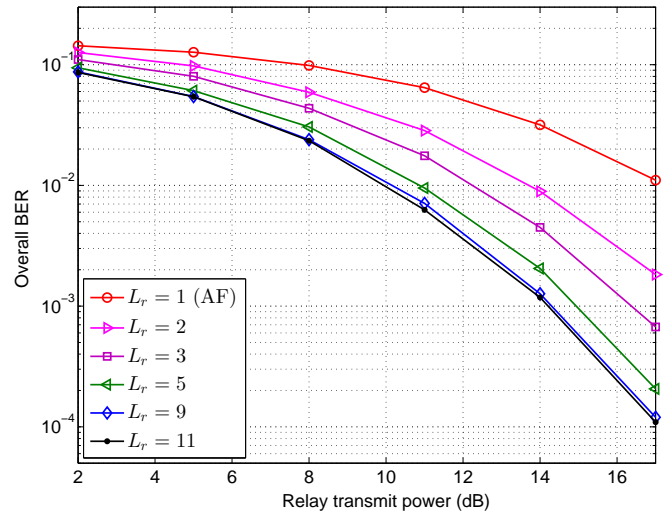}
    \end{psfrags}
} \captionsetup{justification=centering} \caption{Overall BER versus FF relay
transmit power.} \label{fig:BER}
\end{figure}

\begin{figure}[http] \centerline{
    \begin{psfrags}
      \scalefig{0.6}\epsfbox{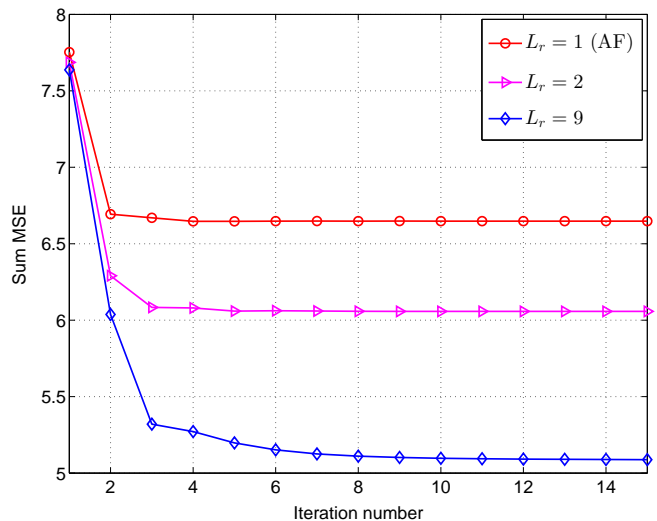}
    \end{psfrags}
} \captionsetup{justification=centering} \caption{Sum MSE versus the number of
iteration.} \label{fig:Converge}
\end{figure}

Next, we investigated the BER performance corresponding to Fig.
\ref{fig:Sum_MSE}. Here,  we assumed uncoded QPSK modulation for
each subcarrier channel. From the result of Fig.
\ref{fig:Sum_MSE}, we obtained the SNR of each subcarrier channel
of the total $N=16$ subcarrier channels for the designed FF relay
filter, transmit filer, receive filter and source power
allocation. Based on this, we  computed the subcarrier BER based
on the  SNR of each subcarrier and averaged all the subcarrier
channel BERs to obtain the overall BER, and the result is shown in
Fig. \ref{fig:BER}. It is seen in Fig. \ref{fig:BER} that the FF
relay significantly improves the BER performance over the  AF
relay. Next, we tested the convergence property of the proposed
algorithm, and Fig. \ref{fig:Converge} shows the result. It is
seen that  the proposed algorithm converges with a few iterations.

\begin{figure}[http] \centerline{
    \begin{psfrags}
      \scalefig{0.6}\epsfbox{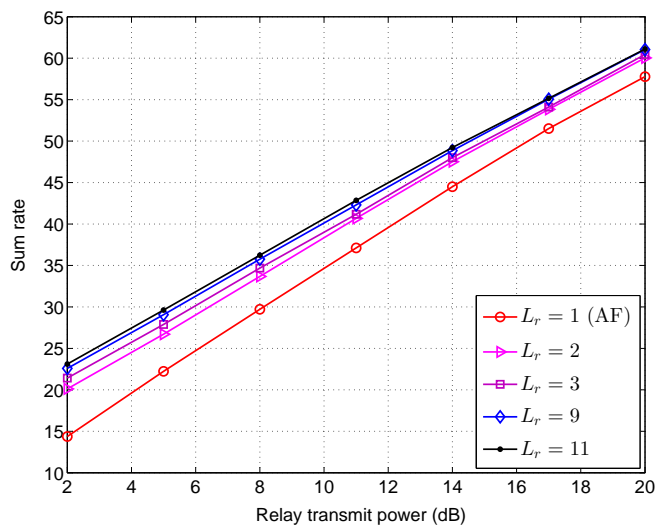}
    \end{psfrags}
} \captionsetup{justification=centering} \caption{Sum rate versus FF relay
transmit power.} \label{fig:rate}
\end{figure}

Finally, we examined the rate performance of the proposed
rate-targeting design method, Algorithm 2. (Rate maximization
 may be the ultimate goal of design in many cases.)
Fig. \ref{fig:rate} shows the result. Again, for the figure 200
channels were randomly realized with $L_f=L_g=3$ and each plotted
value is the average over the 200 channel realizations, and the
sum rate is the sum over the total subcarrier channels. It is
shown in Fig. \ref{fig:rate} that the FF relay improves the rate
performance as the FF relay filter length increases, and the
improvement gap shows that it is worth considering FF relays over
simple AF relays even though FF relays require more processing
than AF relays.

\section{Conclusion}    \label{sec:conclusion}


In this paper, we have considered the joint design of the linear
transceiver and the FF relay for MIMO-OFDM systems for weighted
sum MSE minimization and sum rate maximization, and have proposed
algorithms for this purpose based on alternating optimization that
iterates between optimal design of the FF relay for a MIMO
transceiver at the source and the destination and optimal design
of the MIMO transceiver for a given FF relay filter. We have shown
that  the FF relay design problem for a given MIMO transceiver
reduces to a quadratically constrained quadratic program (QCQP)
and have proposed a solution to this QCQP problem based on
conversion to a semi-definite program (SDP). We have provided some
numerical results to evaluate the performance gain of the FF
relaying scheme over the simple AF scheme for MIMO-OFDM systems.
Numerical results show the effectiveness of the proposed FF relay
design and suggest that it is worth considering the FF relaying
scheme over the widely-considered simple AF scheme for MIMO-ODFM
systems.


%




\newpage
\appendices
\section{$\Ebf_1$ and $\Ebf_2$ matrices} \label{sec:appendix1}

$\Ebf_1$ and $\Ebf_2$  are  ${M_tL_rM_r ~\times~ M_t(N+L_r+L_g-2)(N+L_g-1) M_r}$ matrices and defined as follows:

{ \small
\begin{eqnarray}
&&\Ebf_1= \nonumber \\
&&
\left[
\underbrace{\left|
               \begin{array}{c} \Ibf\\   \bf{0}\\ \vdots\\ \vdots\\ \bf{0}\\ \end{array}
    \underbrace{\begin{array}{c} \bf{0}\\ \vdots\\ \vdots\\ \vdots\\ \bf{0}\\ \end{array} }_{N+L_g-2}
    \left|
                \begin{array}{c} \bf{0}\\ \Ibf\\   \bf{0}\\ \vdots\\ \bf{0}\\ \end{array}
    \underbrace{\begin{array}{c} \bf{0}\\ \vdots\\ \vdots\\ \vdots\\ \bf{0}\\ \end{array} }_{N+L_g-2}
    \right|
    \begin{array}{c}    \cdots\\    \end{array}
    \left|
                \begin{array}{c} \bf{0}\\ \vdots\\ \vdots\\ \bf{0}\\ \Ibf\\   \end{array}
    \underbrace{\begin{array}{c} \bf{0}\\ \vdots\\ \vdots\\ \vdots\\ \bf{0}\\ \end{array} }_{N+L_g-2}
    \right.
\right|}_{M_t(N+L_r+L_g-2)}
\right.
\underbrace{\left|
    \underbrace{\begin{array}{c} \bf{0}\\ \vdots\\ \vdots\\ \vdots\\ \bf{0}\\ \end{array} }_{1}
                \begin{array}{c} \Ibf\\   \bf{0}\\ \vdots\\ \vdots\\ \bf{0}\\ \end{array}
    \underbrace{\begin{array}{c} \bf{0}\\ \vdots\\ \vdots\\ \vdots\\ \bf{0}\\ \end{array} }_{N+L_g-3}
    \left|
    \underbrace{\begin{array}{c} \bf{0}\\ \vdots\\ \vdots\\ \vdots\\ \bf{0}\\ \end{array} }_{1}
                \begin{array}{c} \bf{0}\\ \Ibf\\   \bf{0}\\ \vdots\\ \bf{0}\\ \end{array}
    \underbrace{\begin{array}{c} \bf{0}\\ \vdots\\ \vdots\\ \vdots\\ \bf{0}\\ \end{array} }_{N+L_g-3}
    \right|
    \begin{array}{c}    \cdots\\    \end{array}
    \left|
    \underbrace{\begin{array}{c} \bf{0}\\ \vdots\\ \vdots\\ \vdots\\ \bf{0}\\ \end{array} }_{1}
                \begin{array}{c} \bf{0}\\ \vdots\\ \vdots\\ \bf{0}\\ \Ibf\\   \end{array}
    \underbrace{\begin{array}{c} \bf{0}\\ \vdots\\ \vdots\\ \vdots\\ \bf{0}\\ \end{array} }_{N+L_g-3}
    \right.
\right|}_{M_t(N+L_r+L_g-2)}
\nonumber \\ &&
    \begin{array}{c}    \cdots\\    \end{array} {
\left.\underbrace{\left|
    \underbrace{\begin{array}{c} \bf{0}\\ \vdots\\ \vdots\\ \vdots\\ \bf{0}\\ \end{array} }_{N+L_g-2}
                \begin{array}{c} \Ibf\\   \bf{0}\\ \vdots\\ \vdots\\ \bf{0}\\ \end{array}
    \left|
    \underbrace{\begin{array}{c} \bf{0}\\ \vdots\\ \vdots\\ \vdots\\ \bf{0}\\ \end{array} }_{N+L_g-2}
                \begin{array}{c} \bf{0}\\ \Ibf\\   \bf{0}\\ \vdots\\ \bf{0}\\ \end{array}
    \right|
    \begin{array}{c}    \cdots\\    \end{array}
    \left|
    \underbrace{\begin{array}{c} \bf{0}\\ \vdots\\ \vdots\\ \vdots\\ \bf{0}\\ \end{array} }_{N+L_g-2}
                \begin{array}{c} \bf{0}\\ \vdots\\ \vdots\\ \bf{0}\\ \Ibf\\   \end{array}
    \right.
\right|}_{M_t(N+L_r+L_g-2)}  \right] \otimes\Ibf_{M_r}   }
\end{eqnarray}
}
where $\Ibf=\Ibf_{L_r}$.
\newpage

{\small
\begin{eqnarray}
&&\Ebf_2 = \nonumber \\
&&\left[
    \begin{array}{|c|}
        \Ec_1\\  \bf{0}\\ \vdots\\ \vdots\\ \bf{0}\\ \hline
                \vdots\\ \hline
        \Ec_{M_t}\\  \bf{0}\\ \vdots\\ \vdots\\ \bf{0}\\
    \end{array}
    \begin{array}{c|}
        \Ec_{M_t+1}\\ \Ec_1\\   \bf{0}\\ \vdots\\ \bf{0}\\ \hline
               \vdots\\ \hline
        \Ec_{M_t+M_t}\\ \Ec_{M_t}\\   \bf{0}\\ \vdots\\ \bf{0}\\
    \end{array}
    \begin{array}{c}    \cdots\\    \end{array}
\underbrace{
    \begin{array}{|c|}
        \Ec_{(L_r-1)M_t+1}\\ \vdots\\   \vdots\\ \vdots\\ \Ec_1\\ \hline
                \vdots\\ \hline
        \Ec_{(L_r-1)M_t+M_t}\\ \vdots\\   \vdots\\ \vdots\\ \Ec_{M_t}\\
    \end{array}
}_{\mbox{$L_r$-th block}}
    \begin{array}{c|}
        \Ec_{(L_r)M_t+1}\\ \vdots\\   \vdots\\ \vdots\\ \Ec_{M_t+1}\\ \hline
                \vdots\\ \hline
        \Ec_{(L_r)M_t+M_t}\\ \vdots\\   \vdots\\ \vdots\\ \Ec_{M_t+M_t}\\
    \end{array}
    \begin{array}{c}    \cdots\\    \end{array}
\underbrace{
    \begin{array}{|c|}
        \Ec_{(N+L_g-2)M_t+1}\\ \vdots\\   \vdots\\ \vdots\\ \Ec_{(N+L_g-L_r-1)M_t+1}\\ \hline
                \vdots\\ \hline
        \Ec_{(N+L_g-2)M_t+M_t}\\ \vdots\\   \vdots\\ \vdots\\ \Ec_{(N+L_g-L_r-1)M_t+M_t}\\
    \end{array}
}_{\mbox{$(N+L_g-1)$-th block}}
\underbrace{
    \begin{array}{c|}
        \bf{0}\\    \Ec_{(N+L_g-2)M_t+1}\\  \vdots\\ \vdots\\ \Ec_{(N+L_g-L_r)M_t+1}\\ \hline
               \vdots\\ \hline
        \bf{0}\\    \Ec_{(N+L_g-2)M_t+M_t}\\ \vdots\\ \vdots\\ \Ec_{(N+L_g-L_r)M_t+M_t}\\
    \end{array}
}_{\mbox{$(N+L_g)$-th block}}\right. \nonumber\\
&& \left.
    \begin{array}{c}    \cdots\\    \end{array}
\underbrace{
    \begin{array}{|c|}
        \bf{0}\\     \vdots\\ \vdots\\ \bf{0}\\ \Ec_{(N+L_g-2)M_t+1}\\ \hline
                \vdots\\ \hline
        \bf{0}\\     \vdots\\ \vdots\\ \bf{0}\\ \Ec_{(N+L_g-2)M_t+M_t}\\
    \end{array}
}_{\mbox{$(N+L_g+L_r-2)$-th block}}
\right] ~~\text{and}~~
\Ec_k=
    \left[
    \begin{array}{c}
        e_k^T\\ e_{(N+L_g-1)M_t+k}^T\\ e_{2(N+L_g-1)M_t+k}^T\\ \vdots\\ e_{(M_r-1)(N+L_g-1)M_t+k}^T\\
    \end{array}
    \right]
\end{eqnarray}
}
where $e_i^T$ is the $i$-th row of  $\Ibf_{(N+L_g-1)M_tM_r}$.

\section{} \label{sec:appendix2}
{\it{Proof of Proposition 1 }}

The Lagrangian of \eqref{problem:sourcePA} is given by
    \begin{align}
    {\cal{L}}(p_n[k], \mu)  &= \sum_{n=0}^{N-1}\sum_{k=1}^B  \theta_{nk} (d_n[k]^2 p_n[k]^2 - 2 d_n[k]  p_n[k] + c_n[k] ) + \mu (\sum_{n=0}^{N-1}\sum_{k=1}^B p_n[k]^2 -P_{s,max}) \nonumber \\
    & ~~ - \sum_{n=0}^{N-1}\sum_{k=1}^B\lambda_{n,k} p_n[k]
    \end{align}
where  $\mu \in {\cal{R}}$ and $\lambda_{n,k} \ge 0$ are dual variables associated with the source power constraint and the positiveness of power, respectively.

\noindent Then, the following KKT conditions are necessary and sufficient for optimality because the problem \eqref{problem:sourcePA} is a
convex optimization problem:
\begin{align}
&p_n[k] \ge 0,~~ \sum_{n=0}^{N-1}\sum_{k=1}^B p_n[k]^2 -P_{s,max} = 0, \label{eq:primal} \\
&\mu \in {\cal{R}},~~ \lambda_{n,k} \ge 0,  \label{eq:dual}\\
& \lambda_{n,k} p_n[k]  = 0  \label{eq:complementaryslackness}\\
& \nabla_{p_n[k]} {\cal{L}} = 2\theta_{nk}d_n[k]^2 p_n[k] - 2 \theta_{nk}d_n[k] + 2\mu  p_n[k]  - \lambda_{n,k} = 0 \label{eq:stationary}
\end{align}
for $ n = 0, 1, \cdots, N-1 $ and $k=1, \cdots, B$.

\noindent The gradient \eqref{eq:stationary} can be rewritten as $\lambda_{n,k} = 2(\theta_{nk}d_n[k]^2 + \mu )p_n[k] - 2\theta_{nk} d_n[k] $. Plugging this into  \eqref{eq:dual} and \eqref{eq:complementaryslackness}, we get
\begin{align}
&  \mu p_n[k] \ge \theta_{nk}d_n[k] - \theta_{nk}d_n[k]^2 p_n[k]        \label{eq:condition1} \\
& ((\theta_{nk}d_n[k]^2 + \mu )p_n[k] -  \theta_{nk}d_n[k]) p_n[k] = 0    \label{eq:condition2}
\end{align}
Let us consider the case that $p_n[k] = 0$. Then, \eqref{eq:condition1} is satisfied only if $d_n[k] = 0$ because $d_n[k] \ge 0$.
If $p_n[k] > 0$, $p_n[k] = \left (\frac{\theta_{nk}d_n[k]}{\theta_{nk}d_n[k]^2 + \mu} \right)$ by the complementary slackness \eqref{eq:condition2}. This also satisfies
\eqref{eq:condition1}. Therefore, we get the desired result satisfying the primal constraints \eqref{eq:primal}.
~~~~~~$\hfill{\square}$



\newpage

\vspace{10cm}



\end{document}